\documentclass[cernpreprint=true, PAPER, atlasdraft=false, texlive=2023, UKenglish, texmf, orcidlogo]{atlasdoc}
\usepackage{atlaspackage}
\usepackage{atlasbiblatex}

\usepackage{atlasphysics}
\ifthenelse{\boolean{AtlasDraft}}{\usepackage[output=true, shift=true]}{}

\graphicspath{{logos/}{figures/}}

\usepackage{ANA-STDM-2018-31-PAPER-defs}

\usepackage{multirow}
\AtlasTitle{Fiducial and differential cross-section measurements of electroweak $W\gamma jj$ production in $pp$ collisions at $\sqrt{s} = 13$ TeV with the ATLAS detector}

\AtlasVersion{3.0}

\AtlasAbstract{%
The observation of the electroweak production of a $W$ boson and a photon in association with two jets, using $pp$ collision data at the Large Hadron Collider at a centre of mass energy of $\sqrt{s}=13$~TeV, is reported. The data were recorded by the ATLAS experiment from 2015 to 2018 and correspond to an integrated luminosity of 140~fb$^{-1}$.
This process is sensitive to the quartic gauge boson couplings via the vector boson scattering mechanism and provides a stringent test of the electroweak sector of the Standard Model. Events are selected if they contain one electron or muon, missing transverse momentum, at least one photon, and two jets.
Multivariate techniques are used to distinguish the electroweak $W\gamma jj$ process from irreducible background processes.
The observed significance of the electroweak $W\gamma jj$ process is well above six standard deviations, compared to an expected significance of 6.3 standard deviations.
Fiducial and differential cross sections are measured in a fiducial phase space close to the detector acceptance, which are in reasonable agreement with leading order Standard Model predictions from \textsc{MadGraph5+Pythia8} and \textsc{Sherpa}. The results are used to constrain new physics effects in the context of an effective field theory.

}

\AtlasRefCode{STDM-2018-31}

\PreprintIdNumber{CERN-EP-2024-048}

\AtlasJournal{EPJC}
\AtlasJournalRef{\EPJC 84 (2024) 1064}
\AtlasDOI{10.1140/epjc/s10052-024-13311-6}

\hypersetup{pdftitle={ATLAS document},pdfauthor={The ATLAS Collaboration}}

\begin{document}

\maketitle

\tableofcontents

\section{Introduction} %
%
%

The scattering of two vector bosons, e.g. $WZ\rightarrow W\gamma$, is sensitive to both the triple and quartic electroweak-boson self-interactions~\cite{Eboli:2003nq,PhysRevD.93.093013}. In proton--proton ($pp$) collisions, the $W\gamma jj$ final state can be produced via many different mechanisms, as shown in Figure~\ref{fig:diag_vbswg}.
Electroweak (EW) $W\gamma jj$ production concerns exclusively electroweak interactions of order $\alpha_\text{EW}^4$ at tree level~\cite{PhysRevD.74.073010}, where $\alpha_\text{EW}$ is the electroweak coupling constant.
Although the contributions of interest are vector boson scattering (VBS) interactions involving quartic gauge couplings (QGCs), these cannot be distinguished from other electroweak contributions in a gauge-invariant manner. Thus, the signal process studied in this paper is the combination of all processes of order $\alpha_\text{EW}^4$ shown in Figures~\ref{fig:diag_vbswg}\subref{subfig:EW1}-\ref{fig:diag_vbswg}\subref{subfig:EW3}. The dominant background for \ewkwyjj production concerns processes of order $\alpha_\text{S}^2\alpha_\text{EW}^2$ at tree level in Figure~\ref{fig:diag_vbswg}\subref{subfig:QCD}, where $\alpha_\text{S}$ is the strong coupling constant and the jets are produced via strong interaction vertices; these processes are collectively referred to as \qcdwyjj production in this paper.
Triboson diagrams, such as the one shown in Figure~\ref{fig:diag_vbswg}\subref{subfig:EW2}, do contribute to \ewkwyjj production. However, due to their distinct topology they require separate signal selection and background estimation methods and are thus generally the subject of separate studies.

The large cross-section of \ewkwyjj production predicted by the SM allows differential cross-sections to be measured with higher precision than other
VBS processes. The differential measurements
further enhance our sensitivity to potential anomalous quartic couplings of $WW\gamma\gamma$ and $WW\gamma Z$. In addition to the VBS topology characterised by the two energetic jets in the forward
and backward region, the leptonic decay channel of the \ewkwyjj production has a clean signature in the detector with exactly one charged lepton, missing transverse momentum, and at least one photon.
This paper reports fiducial and differential cross-section measurements in the \ewkwyjj final state using 140 \ifb\ of data recorded between 2015 and 2018 with the ATLAS detector.

\begin{figure}[!htb]
\centering
\subfloat[]{\includegraphics[width=0.4\textwidth]{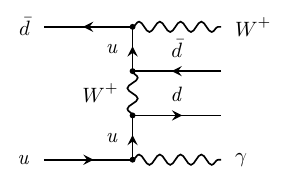}\label{subfig:EW1}}
\subfloat[]{\includegraphics[width=0.4\textwidth]{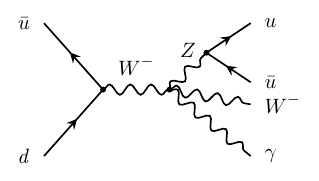}\label{subfig:EW2}}\\
\subfloat[]{\includegraphics[width=0.4\textwidth]{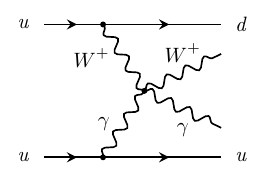}\label{subfig:EW3}}
\subfloat[]{\includegraphics[width=0.4\textwidth]{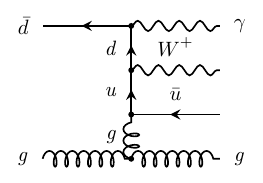}\label{subfig:QCD}}\\
\caption{Representative Feynman diagrams for the $W\gamma jj$ final state: \protect\subref{subfig:EW1} EW $W\gamma jj$ production involving no gauge boson self-interactions, \protect\subref{subfig:EW2} Bremsstrahlung EW $W\gamma jj$ non-VBS production involving quartic gauge boson interactions, \protect\subref{subfig:EW3} $W\gamma$ VBS involving quartic gauge boson interactions, and \protect\subref{subfig:QCD} \QCDwyjj production.}
\label{fig:diag_vbswg}
\end{figure}

The CMS Collaboration reported the observation of the \ewkwyjj process~\cite{CMS-SMP-19-008} and measured \ewkwyjj differential cross-sections~\cite{CMS-SMP-21-011}. This analysis exploits two innovations compared to the CMS publication. First, multivariate techniques involving a neural network are used to isolate the \ewkwyjj signal from the \qcdwyjj background in a VBS-enhanced phase space; the resulting observed yield is corrected for detector effects and a fiducial cross-section is reported in a fiducial phase space at particle level as close as possible to the reconstruction level.
Secondly, observables sensitive to the charge conjugation (C) and parity (P) structure of $WW\gamma\gamma$ and $WW\gamma Z$ couplings are measured. These
measurements can be used to explore new sources of CP violation (CPV) in the gauge-boson sector that may partly explain the predominance of matter over anti-matter in the universe.
The differential cross-sections for \ewkwyjj production are measured as functions of two types of observables: \textit{VBS observables} and \textit{charge conjugation and parity observables}.
\begin{itemize}
\item The VBS observables are those that are used to characterise specifically the VBS nature of events; they are additionally sensitive to anomalous QGCs (aQGCs) via dimension-8 operators of an effective field theory (EFT), which can be parameterised in the \'{E}boli basis~\cite{PhysRevD.93.093013}. These observables include the invariant mass of the dijet system, $m_{jj}$, the transverse momentum of the two jets, $\pt^{jj}$, the lepton transverse momentum, $\pt^l$, and the invariant mass of the lepton and the photon, $m_{l\gamma}$.
\item The charge conjugation and parity observables probe the CP structure of $WW\gamma\gamma$ and $WW\gamma Z$ couplings. Two observables are studied: the signed azimuthal angle difference between the two jets,  $\Delta \phi_{jj}=\phi_f^j-\phi_b^j$, where the two highest transverse-momentum jets are ordered by rapidity $y$ such that $y_f^j>y_b^j$, and the signed azimuthal angle difference between the lepton and the photon, $\Delta\phi_{l\gamma}=\phi_f-\phi_b$, where the lepton and the photon are ordered such that $y_f>y_b$. Measurements of $\Delta \phi_{jj}$ have previously been proposed to constrain the
CP-odd component involving Higgs boson couplings in Vector Boson Fusion (VBF) processes~\cite{HiggsCPPRL} and the $t\bar{t}H$ process~\cite{Higgs2jetCP}.

\end{itemize}

The paper is organised as follows.
Section~\ref{sec:detector} briefly describes the ATLAS detector.
Section~\ref{sec:simulation} documents the simulated signal and background samples used in this analysis.
The object reconstruction and event selection is presented in Section~\ref{sec:reconstruction}.
The background estimate is detailed in Section~\ref{sec:bkg}.
The methodologies used for the fiducial and differential cross-section measurements are discussed in Section~\ref{sec:signal}.
Corrections for detector effects are described in Section~\ref{sec:unfolding}, followed by a description of the systematic uncertainties in Section~\ref{sec:systs}.
The fiducial and differential \ewkwyjj cross-sections are presented in Section~\ref{sec:result}.
Section~\ref{sec:EFTs} presents constraints on the Wilson coefficients of dimension-8 EFT.


%
\section{ATLAS detector} %
\label{sec:detector}
%

\newcommand{\AtlasCoordFootnote}{ ATLAS uses a right-handed coordinate system with its origin at the nominal interaction point (IP) in the centre of the detector and the \(z\)-axis along the beam pipe. The \(x\)-axis points from the IP to the centre of the LHC ring, and the \(y\)-axis points upwards. Polar coordinates \((r,\phi)\) are used in the transverse plane, \(\phi\) being the azimuthal angle around the \(z\)-axis. The pseudorapidity is defined in terms of the polar angle \(\theta\) as \(\eta = -\ln \tan(\theta/2)\) and is equal to the rapidity $ y = \frac{1}{2} \ln \left( \frac{E + p_z c}{E - p_z c} \right) $ in the relativistic limit. Angular distance is measured in units of \(\Delta R \equiv \sqrt{(\Delta y)^{2} + (\Delta\phi)^{2}}\).}

The ATLAS detector~\cite{PERF-2007-01} at the LHC covers nearly the entire solid angle around the collision point.\footnote{\AtlasCoordFootnote}
It consists of an inner tracking detector surrounded by a thin superconducting solenoid, electromagnetic and hadronic calorimeters,
and a muon spectrometer incorporating three large superconducting air-core toroidal magnets.

The inner-detector system (ID) is immersed in a \qty{2}{\tesla} axial magnetic field
and provides charged-particle tracking in the range \(|\eta| < 2.5\).
The high-granularity silicon pixel detector covers the vertex region and typically provides four measurements per track,
the first hit generally being in the insertable B-layer (IBL) installed before Run~2~\cite{ATLAS-TDR-19,PIX-2018-001}.
It is followed by the SemiConductor Tracker (SCT), which usually provides eight measurements per track.
These silicon detectors are complemented by the transition radiation tracker (TRT),
which enables radially extended track reconstruction up to \(|\eta| = 2.0\).
The TRT also provides electron identification information
based on the fraction of hits (typically 30 in total) above a higher energy-deposit threshold corresponding to transition radiation.

The calorimeter system covers the pseudorapidity range \(|\eta| < 4.9\).
Within the region \(|\eta|< 3.2\), electromagnetic calorimetry is provided by barrel and
endcap high-granularity lead/liquid-argon (LAr) calorimeters,
with an additional thin LAr presampler covering \(|\eta| < 1.8\)
to correct for energy loss in material upstream of the calorimeters.
Hadronic calorimetry is provided by the steel/scintillator-tile calorimeter,
segmented into three barrel structures within \(|\eta| < 1.7\), and two copper/LAr hadronic endcap calorimeters.
The solid angle coverage is completed with forward copper/LAr and tungsten/LAr calorimeter modules
optimised for electromagnetic and hadronic energy measurements respectively.

The muon spectrometer (MS) comprises separate trigger and
high-precision tracking chambers measuring the deflection of muons in a magnetic field generated by the superconducting air-core toroidal magnets.
The field integral of the toroids ranges between \num{2.0} and \qty{6.0}{\tesla\metre}
across most of the detector.
Three layers of precision chambers, each consisting of layers of monitored drift tubes, cover the region \(|\eta| < 2.7\),
complemented by cathode-strip chambers in the forward region, where the background is highest.
The muon trigger system covers the range \(|\eta| < 2.4\) with resistive-plate chambers in the barrel, and thin-gap chambers in the endcap regions.

The luminosity is measured mainly by the LUCID--2~\cite{LUCID2} detector that records Cherenkov light produced in the quartz windows of photomultipliers located close to the beampipe.

Events are selected by the first-level trigger system implemented in custom hardware,
followed by selections made by algorithms implemented in software in the high-level trigger~\cite{TRIG-2016-01}.
The first-level trigger accepts events from the \qty{40}{\MHz} bunch crossings at a rate below \qty{100}{\kHz},
which the high-level trigger further reduces in order to record complete events to disk at about \qty{1}{\kHz}.

A software suite~\cite{ATL-SOFT-PUB-2021-001} is used in data simulation, in the reconstruction
and analysis of real and simulated data, in detector operations, and in the trigger and data acquisition
systems of the experiment.


%
\section{Monte Carlo event simulation} %
\label{sec:simulation}
%
%

%
%

%
%

Monte Carlo (MC) event generators are used in the analysis to simulate signal and background events produced in $pp$ collisions. These simulated samples are used to design and optimise the analysis, evaluate systematic uncertainties, and characterise the effects of detector inefficiency and resolution.

Electroweak $W\gamma jj$ production is simulated with the \SHERPA[2.2.12]~\cite{Bothmann:2019yzt}
generator. Matrix elements at leading order (LO) in QCD with up to one additional emission are matched to a parton
shower based on Catani--Seymour dipole factorisation~\cite{Gleisberg:2008fv,Schumann:2007mg} using the MEPS@LO
prescription~\cite{Hoeche:2011fd,Hoeche:2012yf,Catani:2001cc,Hoeche:2009rj}.
Samples are simulated using the \NNPDF[3.0nnlo] parton distribution function (PDF) set~\cite{Ball:2014uwa},
along with the dedicated set of tuned parton-shower parameters
developed by the \SHERPA authors. An alternative \ewkwyjj signal sample is produced using \MGNLO~\cite{Alwall:2014hca} at LO accuracy with the default dynamical scale choice and the \NNPDF[3.1lo] PDF set. \Pythia[8.240] with the dipole recoil option turned on is used to add parton showering, hadronisation, and underlying-event activity.  The A14 ~\cite{ATL-PHYS-PUB-2014-021} set of tuned parton-shower parameters is used for \Pythia[8.240], and \EVTGEN~\cite{Lange:2001uf} is used for the properties of bottom and charmed hadron decays. This alternative \ewkwyjj sample is used to evaluate a systematic uncertainty of the signal production due to the choice of event generator.

The dominant background process, \qcdwyjj, is simulated using \SHERPA[2.2.11]~\cite{Bothmann:2019yzt}. Matrix elements at next-to leading order (NLO) QCD accuracy for up to one additional parton and LO accuracy for up to three additional parton emissions are matched and merged with
the \SHERPA parton shower based on Catani--Seymour
dipole factorisation~\cite{Gleisberg:2008fv,Schumann:2007mg} using the MEPS@NLO
prescription~\cite{Hoeche:2011fd,Hoeche:2012yf,Catani:2001cc,Hoeche:2009rj}. This sample uses \NNPDF[3.0nnlo] for the matrix element calculation with default parameters for parton showering, hadronisation, and underlying-event activity.
An alternative \qcdwyjj sample is produced using \MGNLO at NLO accuracy for events with up to three partons in the final state with the \NNPDF[3.0nlo] PDF set and is interfaced with \Pythia[8.240] with A14 parameters to provide parton showering, hadronisation, and underlying-event activity. This alternative \qcdwyjj sample is used to evaluate a systematic uncertainty of the \qcdwyjj background estimate due to the choice of event generator.

Prompt background events are those that contain at least one prompt photon, exactly one prompt charged lepton, and two jets.
\SHERPA[2.2.12], with the same settings as EW and \qcdwyjj, is also used to simulate EW and strong $Z\gamma jj$ production.  The remaining prompt backgrounds can arise from $tW\gamma$, $tq\gamma$, and $t\bar{t}\gamma$ processes, which are simulated using \MGNLO~\cite{Alwall:2014hca} with \NNPDF[2.3nlo] for $tW\gamma$ and $t\bar{t}\gamma$ at LO, and $tq\gamma$ at NLO. To provide parton showering, hadronisation, and underlying-event activity these samples are interfaced \Pythia[8.235], \Pythia[8.212], and \Pythia[8.240], respectively, with A14 parameters.

Contributions from background events arising from jets misidentified as leptons or photons, non-prompt leptons or photons from decays of hadrons, electrons reconstructed as photons, and photons arising from a separate $pp$ interaction (``\pileup photons'') are estimated by using the data-driven methods that are outlined in Section~\ref{sec:bkg}. The term \textit{fake leptons} is used to collectively refer to either non-prompt leptons or hadronic jets misreconstructed as leptons. MC simulated samples including $W$+jets, $Z$+jets, $tW$, $t\bar{t}$, diboson, and multijet events are used in the validation of data-driven methods and the evaluation of systematic uncertainties of the background estimations. In particular, simulated $W$+jets and $Z$+jets events, which are used in the validation of the non-prompt photon estimate, are produced at NLO accuracy using \SHERPA[2.2.11] with \NNPDF[3.0nnlo] and default parameter tunes for parton showering, hadronisation, and underlying-event activity. Background events from electrons misreconstructed as photons can arise from $tW$, $t\bar{t}$, and $Z$+jets processes. The $tW$ and $t\bar{t}$ processes are simulated using \POWHEG at NLO accuracy with \NNPDF[3.0nnlo] and \Pythia[8.230] with A14 parameters to provide parton showering, hadronisation, and underlying-event activity. Diboson and multijet events are used in the data-driven estimate of fake leptons. The diboson sample is simulated with \SHERPA[2.2.2]
at NLO with \NNPDF[3.0nlo] and default parameter tunes for parton showering, hadronisation, and underlying-event activity. The multijet sample is simulated using \Pythia[8.235] at LO accuracy with \NNPDF[2.3lo] and A14 tunes for parton showering.

All simulated signal and background events are processed through the full ATLAS detector simulation using \geant~\cite{Agostinelli:2002hh} and then reconstructed using the same algorithms as the recorded data events. Differences between reconstructed leptons, photons, jets and missing transverse momentum in simulations and data are corrected using event-by-event scale factors with $\pt$ and $\eta$ dependence. The effect of multiple interactions in the same and neighbouring bunch crossings (\pileup) is modelled by overlaying the simulated hard-scattering event with inelastic \(pp\) events simulated with \PYTHIA[8.186]~\cite{Sjostrand:2007gs} using the \NNPDF[2.3lo] set of PDFs~\cite{Ball:2012cx} and the A3 set of tuned parton-shower parameters~\cite{ATL-PHYS-PUB-2016-017}.

%


%
\section{Object reconstruction and event selection} %
\label{sec:reconstruction}
%

This analysis is performed using $pp$ collision data collected between 2015 and 2018 at a centre-of-mass energy of $\sqrt{s}=13$~\TeV, corresponding to an integrated luminosity of 140 $\ifb$~\cite{Lumi2022:ATLAS}.

Events are required to satisfy the unprescaled single-lepton triggers that have transverse momentum
thresholds of 20--26~\GeV, depending on the charged lepton flavour and data-taking periods~\cite{TRIG-2018-01, TRIG-2018-05}; this is complemented by triggers with higher \pT thresholds and no isolation requirements to increase the overall trigger efficiency. Events are also required to satisfy detector and data quality
requirements during stable beam conditions~\cite{DAPR-2018-01}.
Candidate $pp$ interaction vertices are reconstructed using charged-particle tracks; each candidate vertex must have at least two tracks with $\pt>500$~\MeV~\cite{ATL-PHYS-PUB-2015-026}.
The vertex with the highest scalar sum of track $\pt^2$ is selected as the primary vertex. Events must contain at least one electron or muon, missing transverse momentum, at least one photon, and at least two jets.

Muons are reconstructed by matching tracks in the MS to a corresponding track in the ID\@. Each muon must satisfy the \textit{tight} identification
criteria and \textit{tight} isolation working point~\cite{MUON-2018-03}. Muons are constrained to originate from the primary vertex by requiring $|d_0/\sigma_{d_0}|<3$
and $|z_0\sin\theta|<0.5$~mm, where $d_0$ is the distance of closest approach to the primary vertex in the transverse plane, with an
uncertainty $\sigma_{d_0}$, and $z_0$ is the longitudinal difference between the point at which $d_0$ is defined and the primary vertex. Selected
muons are required to have $\pt>30$~\GeV and $|\eta|<2.5$.

Electrons are reconstructed from clusters of energy deposits in the ECAL and matched to a track reconstructed in the ID\@. Each electron must
satisfy the \textit{tight} identification and \textit{tight} isolation working points~\cite{EGAM-2018-01}.
Electrons are required to have $\pt>30$~\GeV and $|\eta|<2.47$, excluding the calorimeter transition region $1.37<|\eta|<1.52$, $|d_0/\sigma_{d_0}|<5$
and $|z_0\sin\theta|<0.5$~mm.

Photons are also reconstructed from clusters of energy deposits in the ECAL\@. Both converted and unconverted photons are used in the analysis.
Converted photons are defined as photon clusters that are matched to one or two ID tracks consistent with a conversion vertex, while unconverted photons are defined as photon clusters that are matched to neither an ID electron track nor a conversion vertex. Each photon is required to satisfy \textit{tight} identification and \textit{tight} isolation working points~\cite{EGAM-2018-01}.
Photons passing a \textit{loose} identification requirement are retained in the analysis to define control regions for the data-driven background estimation method used to determine the background from jets misidentified as photons, which is described in Section~\ref{sec:bkg}.
The \textit{tight} isolation requirement is defined as $\et^\mathrm{cone40}<0.022\times\pt+2.45$~\GeV, where $\et^\mathrm{cone40}$ is computed as the sum of transverse energies of positive-energy topological clusters in the calorimeter within a distance of $\Delta R=0.4$ around the photon candidate~\cite{EGAM-2018-01}. Photons are required to have $\pt>22$~\GeV and $|\eta|<2.37$, excluding the calorimeter transition region $1.37<|\eta|<1.52$.

Jets are reconstructed using a particle-flow method~\cite{PERF-2015-09}
that combines charged-particle tracks in the ID with topo-clusters formed from energy deposits in the calorimeters.
The anti-$k_t$ algorithm~\cite{Cacciari:2008gp} with radius parameter $R=0.4$ is used to define the jet. Each jet is required to have $\pt>25$~\GeV and $|\eta|<4.4$. Jets with $\pt<60$~\GeV and $|\eta|<2.4$ are required to originate from the primary vertex by using the \textit{tight} working point of the jet vertex tagger (JVT)~\cite{PERF-2014-03}. Jets that originate from noise bursts in the calorimeters are removed~\cite{ATLAS-CONF-2015-029}. Events must have at least two jets with $\pt>50$~\GeV. Jets with $|\eta|<2.5$ that contain $b$-hadrons are identified by a multivariate algorithm (DL1r)~\cite{ATL-PHYS-PUB-2017-013} with 85\% tagging efficiency working point. Events identified as containing a $b$-jet by the DL1r algorithm are excluded to suppress backgrounds from processes involving top quarks.

The magnitude of the missing transverse momentum, \met, and its direction are calculated from the negative vector sum of a track-based soft term and all reconstructed electrons, muons, photons, and jets~\cite{PERF-2016-07}. The soft term is calculated from tracks from the primary vertex that are not matched to any hard physics objects. Events are required to satisfy $\met>30$~\GeV.

To resolve ambiguities in the object reconstruction, an overlap removal procedure is applied. Jet candidates are removed if $\Delta R(j,l)<0.2~(l=e,\mu)$. Then, leptons are removed if $\Delta R(l,j)<0.4$,  and photons are removed if $\Delta R(\gamma,l)<0.4$ or $\Delta R(\gamma,j)<0.4$. Finally, electron candidates are removed if shared ID tracks exist between a muon and an electron.

Events are required to satisfy $\mtw>30 $~\GeV, where $\mtw = \sqrt{2 p^{l}_{\mbox{\footnotesize{T}}} \MET (1-\cos{\Delta \phi})}$ and $\Delta\phi$ is the azimuthal angle difference between the lepton and missing transverse momentum in the transverse plane, relative	 to the beam axis. In addition, to reject events that have topologies consistent with leptonic $Z$ decays in which one charged lepton is reconstructed as a photon, the invariant mass of the lepton and the photon must satisfy $|m_{l\gamma}-m_Z|>10$~\GeV. To further remove events containing two prompt charged leptons consistent with a $Z$ boson or two $W$ bosons, events are rejected if a second lepton satisfying the following requirements is present: $\pt>7$~\GeV, the lepton-specific pseudorapidity requirements, both track impact parameter requirements, and \textit{loose} identification~\cite{MUON-2018-03,EGAM-2018-01}. %
Finally, the two leading jets must have a rapidity difference $\dyjj > 2$ and invariant mass $m_{jj} > 500$~\GeV, which ensures a topology consistent with \ewkwyjj production.

The preceding selection criteria comprise the baseline selection. Baseline selected events are further divided into different signal and control regions, which differ depending on the purpose. The selection for the fiducial cross-section measurement of \ewkwyjj production divides the baseline region into a signal region, \SRnn, and a control region, \CRnn, by counting jets in the rapidity interval between the two leading jets, \Ngapjets. In \ewkwyjj events, the two leading jets are produced via an electroweak interaction and there is thus little hadronic activity between them; hence each event in \SRnn is required to satisfy \Ngapjets = 0, and \CRnn is defined by $\Ngapjets > 0$.

For the differential cross-section measurement, it is additionally required that events have $m_{jj}>1$~\TeV to enhance the \ewkwyjj signal purity; these events are then divided into three control regions (\CRA, \CRB, and \CRC) and one signal region (\SR). The centrality of the lepton-photon system relative to the VBS tagged jets, $j_1$ and $j_2$, is defined as $\lycentrality=|(y_{l\gamma}-(y_{j_1}+y_{j_2})/2)/(y_{j_1}-y_{j_2})|$ and is used to form three control regions (\CRA, \CRB, and \CRC) and one signal region (\SR), where $y_{l\gamma}$ is the rapidity of the lepton-photon system.
In \ewkwyjj events, the $W$ boson and photon are generally produced centrally, between the two leading jets, and thus these events have low $\lycentrality$. The variables $\lycentrality$ and \ngapjets are chosen to define the four regions because they are uncorrelated.
The \SR is defined by requiring that there must be little hadronic activity in the region between the two leading jets and that the reconstructed $l\gamma$ system is produced centrally ($\Ngapjets=0, \lycentrality<0.35$).
The remaining three regions are control regions with small \ewkwyjj contribution, and are used to constrain the dominant background from
\qcdwyjj production: \CRA ($\Ngapjets >0, \lycentrality < 0.35$), \CRB ($\Ngapjets > 0, 0.35 < \lycentrality < 1$), and \CRC ( $\Ngapjets = 0, 0.35 < \lycentrality < 1$).

The signal and control region selection requirements for both the fiducial and differential cross-section measurements are summarised in Table~\ref{tab:sum_SRCR}.

\begin{table}[h!]
\begin{center}
\caption{Summary table for signal and control regions for the fiducial and differential cross-section measurements.}
\begin{tabular}{l|cccc}
\hline\\
Fiducial cross-section &\multicolumn{2}{c}{\raisebox{-0.05ex}{\SRnn}} & \multicolumn{2}{c}{\raisebox{-0.05ex}{\CRnn}}\\[1.0ex]
\hline\\
&  \multicolumn{2}{c}{\raisebox{0.05ex}{$\ngapjets = 0$}} & \multicolumn{2}{c}{\raisebox{0.05ex}{$\ngapjets > 0$}} \\[1.0ex]
\hline
\hline
Differential cross-section & \SR & \CRA & \CRB & \CRC \\
\hline
\rule{0pt}{20pt}\multirow{2}{*}{~~$\mjj>1$~\TeV} 	& $\ngapjets = 0$ & $\ngapjets > 0$ & $\ngapjets > 0$ & $\ngapjets = 0$ \\
& $\lycentrality < 0.35$ & $\lycentrality < 0.35$ & $ 0.35 < \lycentrality< 1$ & $ 0.35 < \lycentrality < 1$ \\
\hline
\end{tabular}
\label{tab:sum_SRCR}
\end{center}
\end{table}

Table~\ref{tab:nn_prefit} shows the number of signal and background events in \SRnn and \CRnn, after computing the data-driven backgrounds, as described in Section~\ref{sec:bkg}. The \qcdwyjj process accounts for 63\% of the event yield in \CRnn and 52\% in the \SRnn. The remaining prompt backgrounds, including top quark processes and EW and strong $Z\gamma jj$ processes, contribute 11\% in \CRnn and 8\% in \SRnn, while the non-prompt background fraction is 22\% in \CRnn and 23\% in \SRnn.

\begin{table}[h!]
\begin{center}
\caption{Expected number of events in the signal and control regions used for the fiducial cross-section measurement and observation of \ewkwyjj production. Statistical and systematic uncertainties estimated in Section~\ref{sec:systs} are included for each component. The number of observed events in each region are included for comparison. The ``non-prompt'' background category includes non-prompt photons and fake leptons.}
\begin{tabular}{lcc}
\hline
\rule{0pt}{10pt} & \SRnn $\left(\ngapjets=0\right)$ &  \CRnn $\left(\ngapjets>0\right)$\\
\hline
\hline
\ewkwyjj   & $520\pm 141$ & $120\pm 49$\\
Strong $W\gamma jj$  & $1550\pm 830$ & $1970\pm 950$\\
Non-prompt    & $692\pm 57$ & $698\pm 58$\\
Top quark processes   & $109\pm 18$ & $183\pm 37$\\
EW + strong $Z\gamma jj$   & $128\pm 34$ & $163\pm 77$\\
\hline
Total   & $3000\pm 830$ & $3140\pm 960$\\
\hline
\hline
Data   & 3341 & 3143 \\
\hline
\end{tabular}
\label{tab:nn_prefit}
\end{center}
\end{table}

%


%
\section{Background estimation} %
\label{sec:bkg}
%

The main source of background arises from \qcdwyjj production. This background is estimated by using MC simulation and constrained using data in the
control regions defined in Table~\ref{tab:sum_SRCR}; a detailed description of this procedure is provided in Section~\ref{sec:signal}.
Additional prompt backgrounds arise from $Z\gamma jj$ production and the production of one or more top quarks in association with photons. These prompt backgrounds are estimated by using MC simulations described in Section~\ref{sec:simulation}.

The largest non-prompt background arises primarily from $W$+jet production, where a jet is misidentified as a photon.
This non-prompt photon background is estimated by using a data-driven template fit to the \pt-corrected photon isolation energy ($\etisoy=\et^\mathrm{cone40}-0.022\times\pt$) distributions of prompt and non-prompt photons.  %
The prompt photon template is determined from prompt photons in $W\gamma jj$ MC simulation. %
A control region enriched with non-prompt photons is defined by requiring photons to satisfy the \textit{loose} identification criteria but fail to satisfy the selection criteria for at least one of four variables in the \textit{tight} identification that define the photon shower shape. This selection is referred to as \textit{LoosePrime4}~\cite{EGAM-2018-01}. The shape of the non-prompt photon template is extracted by parameterising the \etisoy distribution in this control region.
Contributions from processes with prompt photons in the non-prompt photon control region are minimal but are accounted for with a systematic uncertainty of the background estimate.
The fractions of prompt and non-prompt photons are extracted for each bin of each distribution using an unbinned, maximum-likelihood fit of the prompt and non-prompt templates to data.
The extracted yields are verified using an ``ABCD sideband method''~\cite{STDM-2010-08} with the two dimensional plane defined by photon identification and isolation requirements.
The dominant uncertainty in the non-prompt photon background is due to the choice of photon identification criteria used to define the non-prompt template regions. This non-prompt photon definition is varied by requiring photons to fail to satisfy at least one of two, three, or five shower shape selection variables used to define the tight identification~\cite{EGAM-2018-01},
and the systematic uncertainty due to this identification choice varies from 4\% to 16\%.
Other sources of systematic uncertainty arise from statistical uncertainties of the samples in the template and fit regions; prompt photon leakage into the non-prompt template regions, estimated from signal MC simulation; and the effect of photon isolation energy modelling uncertainties on the prompt photon templates.

A ``fake factor'' method is used to estimate the background from fake leptons: charged leptons arising from misreconstructed jets or in-flight decays of hadrons.
The method defines \textit{tight} leptons as those satisfying all selection criteria described in Section~\ref{sec:reconstruction} and \textit{non-tight} leptons as those failing to satisfy either the \textit{tight} identification or \textit{tight} isolation criteria. \textit{Non-tight} leptons must still satisfy \textit{loose} identification in addition to the identification and isolation requirements associated with the single lepton triggers. The fake factor is defined as the ratio of events with one \textit{tight} lepton to those events with one \textit{non-tight} lepton. It is calculated differentially in lepton $\pt$ and $\eta$ in a non-prompt control region defined by exactly one lepton and at least one jet.
Contributions from processes with prompt leptons are subtracted from the data prior to calculating the fake factor.
The non-prompt lepton background is then determined in each bin of each measured distribution by applying the fake factor to prompt, background-subtracted data events that satisfy all the selection criteria except the \textit{tight} lepton criteria.
Sources of uncertainty arise from data and MC sample size in the control regions, relative fractions between heavy flavour decays and photon conversions, control region definitions for fake factor calculation, theoretical uncertainties of the prompt background, and the binning in $\pt^l$ and $\eta$.

An electron can be misreconstructed as a converted photon ($e\rightarrow \gamma$) if the track reconstruction algorithm either fails to associate a B-layer hit to the track or associates a spurious conversion track to the electron. This background arises predominantly from $Z$+jets and $t\bar{t}$ processes, and the fake rate is determined using data by selecting candidate $ee$ and $e\gamma$ events in a range of $\pm20$~\GeV around the $Z$ boson mass. The object selections for the electron and photon candidates satisfies the same criteria specified in Section~\ref{sec:reconstruction}. Compared with the analysis region selections described in Section~\ref{sec:reconstruction}, events must satisfy the following criteria: $\met<20$~\GeV,  $\mtw<20$~\GeV, and there are no jet requirements. The $ee$ and $e\gamma$ event yields are obtained by fitting a double-sided Crystal Ball function~\cite{CrystalBall} to $m_{ee}$ or $m_{e\gamma}$ to model the $Z$ boson decay along with an exponential function to account for backgrounds.
The ratio of $N_{e\gamma}/N_{ee}$ is parameterised as functions of electron $\pt$ and $\eta$ and applied to data events with $eejj$ and $e\mu jj$ final states, where one electron replaces the photon in the event selection, to estimate the $e\rightarrow \gamma$ background in electron and muon channels, respectively.
Sources of systematic uncertainties for the $e\rightarrow \gamma$ fake rate arise from variations in the invariant mass fit range, varying the binning in $\pt$ and $\eta$, and replacing the exponential background parameterisation with a fourth-order Bernstein polynomial~\cite{BernsteinPoly}.

No explicit requirement is imposed on the longitudinal position of the photon relative to the primary vertex, $\Delta z_\gamma$, as it is not well-measured for unconverted photons.
As a result, a combinatorial background arises whereby a photon originating from one $pp$ interaction is selected alongside a $Wjj$ event from another $pp$ interaction from the same bunch crossing.
This pile-up photon background is estimated by using a data-driven method~\cite{STDM-2018-04} that exploits the difference in $\Delta z_\gamma$ between hard scatter and pile-up photons.
Because the \pileup photon fraction, $f_\mathrm{PU}$, is independent of the photon conversion status, events are selected if they contain a converted photon with two tracks and a conversion vertex with radius $<125$~mm such that the conversion occurs within the pixel detector, allowing for an accurate measurement of  $\Delta z_\gamma$.
The sidebands of the $\Delta z_\gamma$ distribution in background-subtracted data and $W\gamma jj$ simulation are used to estimate the fraction of events in each analysis region that have photons originating from a pile-up interaction. A control region dominated by pile-up photon background is defined by $|\Delta z_\gamma|>50$~mm.
Due to the limited number of events in this data control region, the statistical precision of $f_\mathrm{PU}$ is almost $90\%$ while the systematic uncertainty is negligible compared with the statistical uncertainty. The pile-up background contains a component where a pile-up jet can be misidentified as a photon. The fraction of prompt photons is estimated by using inclusive prompt photon events obtained from data recorded using single photon triggers, and $f_\mathrm{PU}$ is then multiplied by this fraction.
The estimated $f_\mathrm{PU}$ is $(1.7\pm1.6)\%$ in \SRnn and $(0.45\pm0.39)\%$ in \CRnn; it is neglected for the differential cross-section measurement due to the limited number of events in the control region of pile-up photons arising from the additional requirement of $\mjj>1$~\TeV. The procedure to extract the \ewkwyjj fiducial cross-section is repeated without the pile-up photon background, and the uncertainty on the fiducial cross-section is unchanged.

%


%
\section{Signal extraction} %
\label{sec:signal}

\subsection{Signal extraction for observation}
\label{subsec:inclusive_signal}
%

A neural network (NN) is used to classify signal and background processes. The NN is trained on the \ewkwyjj process as the signal model and the sum of \qcdwyjj, $Z\gamma jj$, and top quark processes as the background model, which are weighted according to their cross-sections. All events used in the training satisfy the signal region \SRnn selection described in Section~\ref{sec:reconstruction}. Two NN classifiers are created after training on two statistically independent but otherwise identical samples. %
A binary cross-entropy loss function is used during the training and the loss for each event is weighted by its weight.
The NN consists of a batch normalisation layer followed by three densely connected hidden layers of 512 nodes each. Each hidden node uses a LeakyRelu activation function~\cite{leaky_relu}. The last hidden layer is densely connected with the output layer that uses a sigmoid activation function.
The model is trained with the Adam learning rate optimiser~\cite{AdamOptimizer} with an initial learning rate of $3\times10^{-5}$.
The NN is trained to discriminate signal from background using 13 kinematic observables.
These 13 observables, ranked according to importance, consist of the lepton-photon centrality, $\xi_{l\gamma}$; pseudorapidity difference between the two jets, $\Delta \eta_{jj}$; angular distance between the lepton-photon system and the dijet system, $\Delta R(l\gamma,jj)$;  $\pt^{\gamma}$; $\pt^{l}$; photon centrality, $\xi_{\gamma}$; angular distance between leading jet and photon, $\Delta R(j_{\mathrm{lead}},\gamma)$; angular distance between lepton and photon,  $\Delta R(l,\gamma)$;  transverse momentum of the lepton-photon system, $\pt^{l\gamma}$; azimuthal angle difference between the lepton-photon and the dijet system, $\Delta \phi(l\gamma,jj)$; $\mtw$; leading jet transverse momentum, $\pt^{j_{\mathrm{lead}}}$; and $\eta_{j_{\mathrm{lead}}}$.
The optimal hyperparameters are determined using a grid search by repeating the training with different combinations of the number of hidden layers, number of nodes, and learning rate; the model architecture with the lowest loss on test data is chosen.
The NN trained on one statistically independent sample is applied to the other for evaluation.
No significant differences between the two NNs are found, and thus the average value of the two NN output scores is used for data and data-driven background events.

A profile likelihood fit to the NN score in the \SRnn and \CRnn region is performed simultaneously to maximise the likelihood for observing $n^\mathrm{data}$ events given the number of predicted events. The expected number of events depends on two floating normalisation factors: $\mu_\mathrm{strong}$ for \qcdwyjj and $\mu_\mathrm{EW}$ as the  \ewkwyjj signal strength. The likelihood function $\mathcal{L}$ can be written in a simplified form as the product of Poisson distributions multiplied by the product of Gaussian constraints,

\begin{equation}
\mathcal{L} = \prod_{r}\prod_{i} P(n_{r,i}^\mathrm{data}|\mu_\mathrm{EW} S_{r,i}(\boldsymbol{\theta}) + B_{r,i}(\mu_\mathrm{strong},\boldsymbol{\theta}))\times \prod_{j} G(\theta_j),
\end{equation}

where $r$ is either \SRnn or \CRnn, $i$ represents the bin number of the NN score in region r, $\boldsymbol{\theta}$ refers to nuisance parameters constrained by the Gaussian term $G(\theta_j)$ for each systematic uncertainty source $j$, $S_{r,i}(\boldsymbol{\theta})$ is the number of predicted events from \ewkwyjj simulation, and $B_{r,i}(\boldsymbol{\theta})$ is the number of background events.

The shape of the NN score for \ewkwyjj and \qcdwyjj is taken from the corresponding MC while the normalisation for each process is floated,
namely, $\mu_\mathrm{EW}$ and $\mu_\mathrm{strong}$, respectively.
The other backgrounds are allowed to vary within their respective uncertainties. The \ewkwyjj measurement depends on a set of nuisance parameters that represent the impact of uncertainties on the fit; these uncertainties are discussed in Section~\ref{sec:systs}. The observed significance is evaluated by performing a background-only simultaneous fit to data in \SRnn and \CRnn to estimate the probability of rejecting the background-only hypothesis. The signal region was blinded when analysis decisions and optimisations were made to avoid biasing the measurement.

Figure~\ref{fig:observables_postfit} shows the top three ranked observables of the NN in the signal region with the values of nuisance parameters and normalisation factors obtained from the fit of the NN score distributions to templates in different variables results and Figure~\ref{fig:nn_score_postfit} shows the output NN score in \SRnn and \CRnn after performing the profile likelihood fit in the corresponding observables. Good agreement with the observed data is seen in both the regions except for the slight shape difference in $\xi_{l\gamma}$ between MC and data. Instead of injecting the nuisance parameter from the fit to the NN score, the maximum-likelihood fit was performed to $\xi_{l\gamma}$ directly to obtain the nuisance parameters, resulting in excellent agreement between MC and data after the fit. Therefore the maximum-likelihood fit has sufficient flexibility to give reliable results despite mismodelling between MC and data before the fit.

\begin{figure}[!htb]
\centering
\subfloat[]{\includegraphics[width=0.33\textwidth]{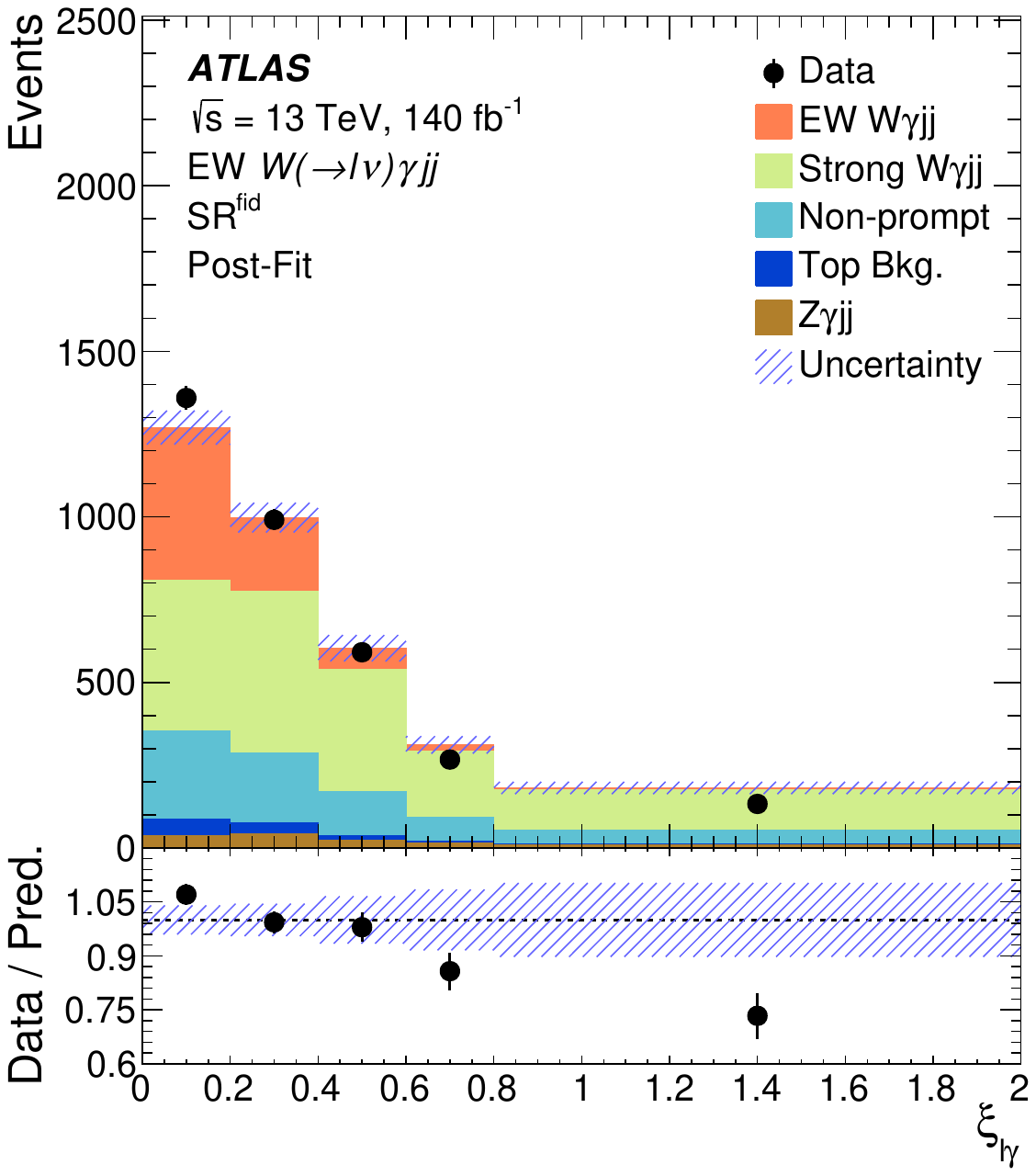}\label{subfig:ly_xi}}
\subfloat[]{\includegraphics[width=0.33\textwidth]{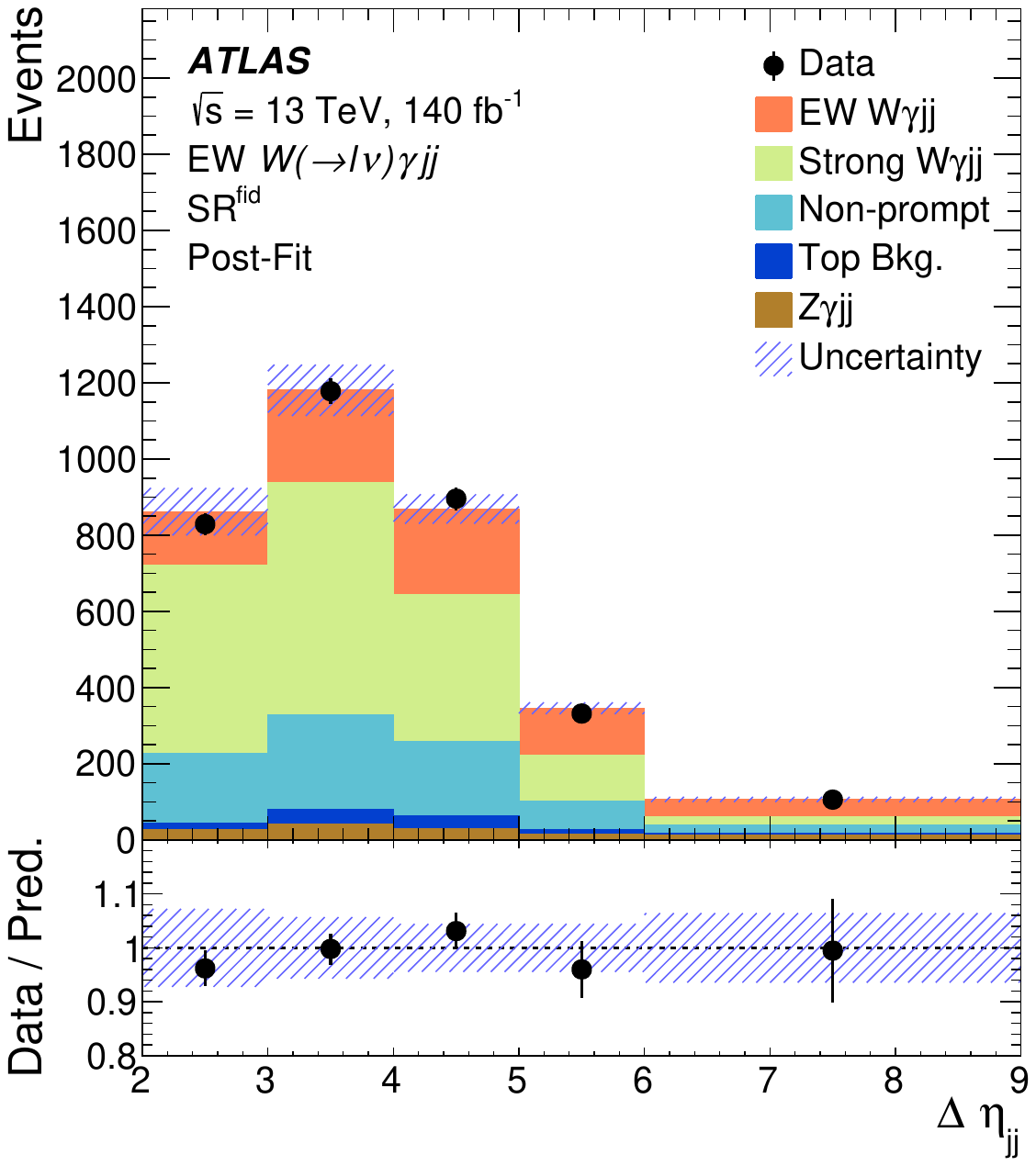}\label{subfig:jj_deta}}
\subfloat[]{\includegraphics[width=0.33\textwidth]{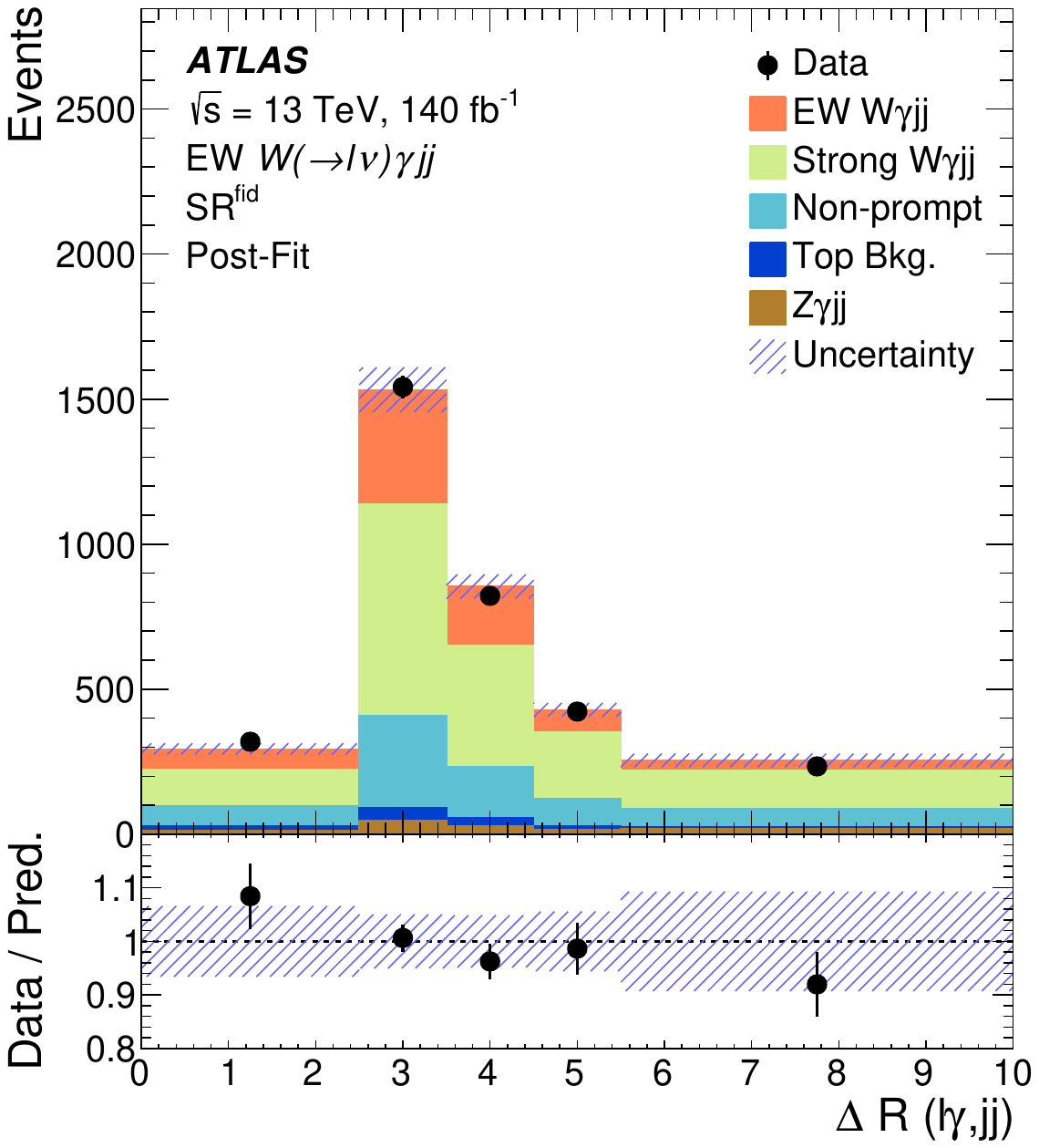}\label{subfig:dR_ly_jj}}
\caption{Distributions of the predicted and observed yields for the three highest ranked variables in the NN after performing the profile likelihood fit, as functions of \protect\subref{subfig:ly_xi} $\xi_{l\gamma}$, \protect\subref{subfig:jj_deta} $\Delta \eta_{jj}$, and \protect\subref{subfig:dR_ly_jj} $\Delta R(l\gamma,jj)$. The observed data is represented by solid circles and the associated vertical error bar represents the statistical uncertainty of the data. The predicted yields comprise simulated \ewkwyjj signal, backgrounds from non-prompt photons and leptons that are estimated by using data-driven methods, and backgrounds that are estimated with simulation. The hashed band represents the quadrature sum of the statistical and systematic uncertainties.}
\label{fig:observables_postfit}
\end{figure}

\begin{figure}[!htb]
\centering
\subfloat[]{\includegraphics[width=0.4\textwidth]{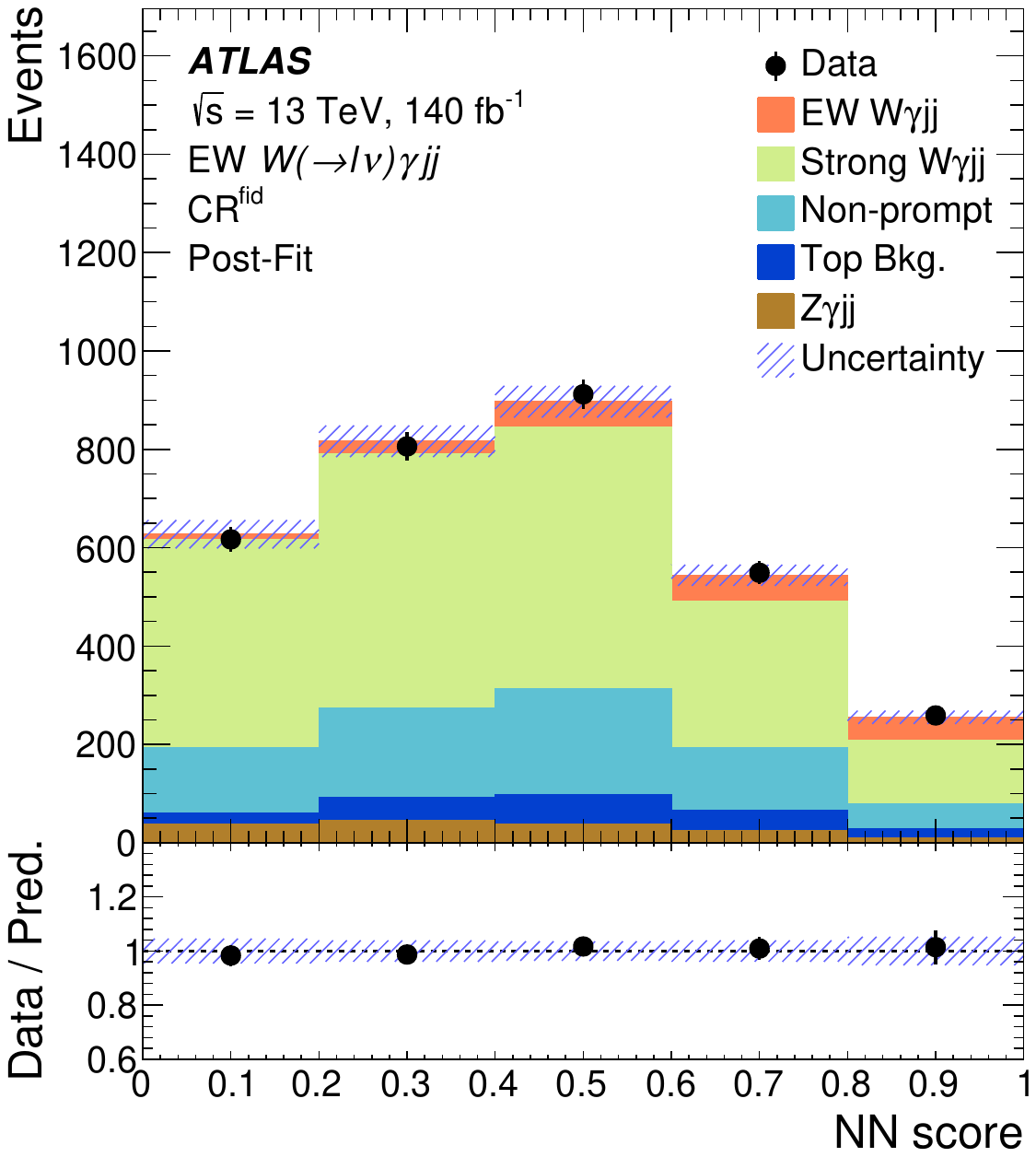}\label{subfig:NN_CR}}
\subfloat[]{\includegraphics[width=0.4\textwidth]{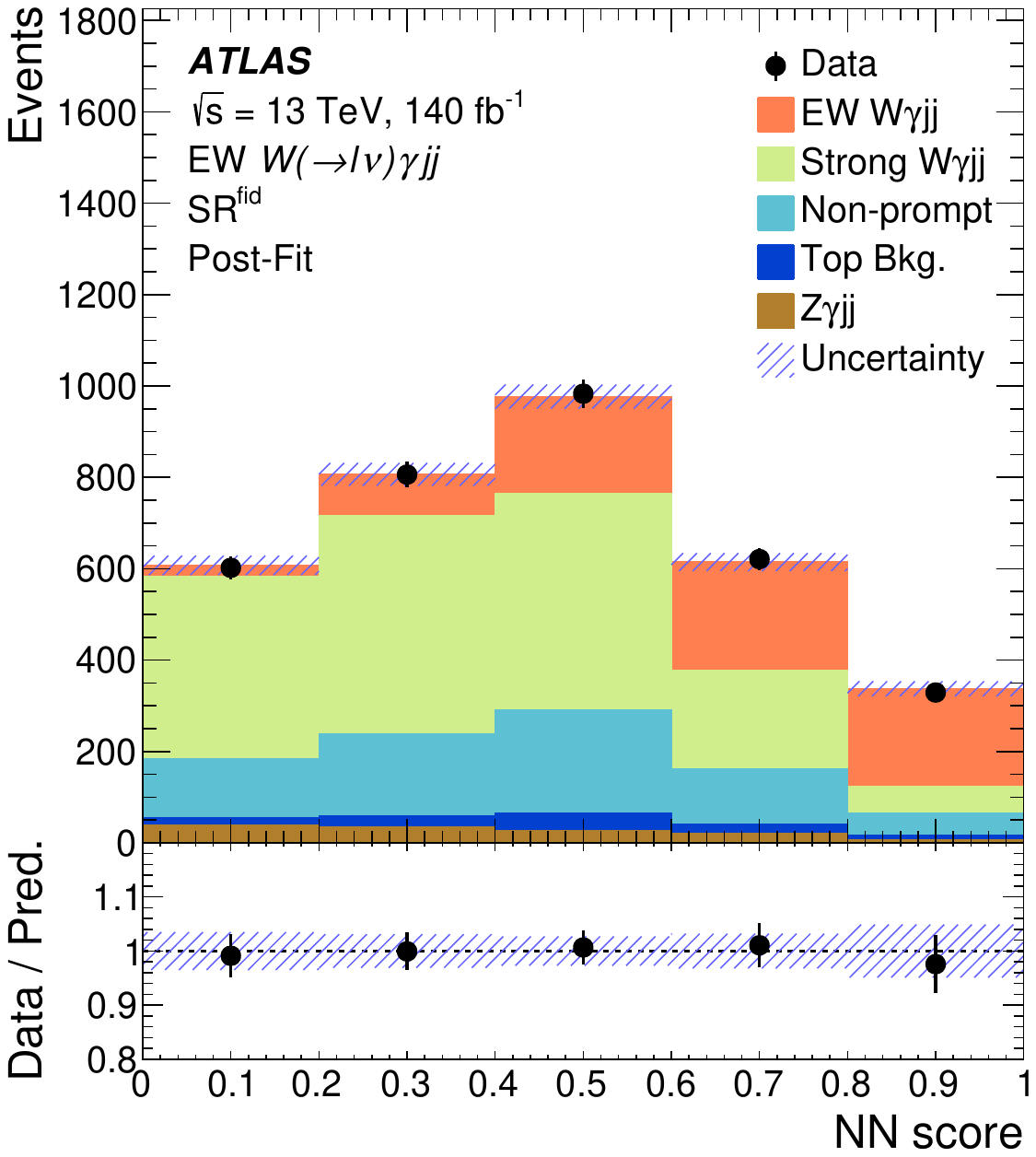}\label{subfig:NN_SR}}
\caption{Distribution of the predicted and observed yields as a function of the NN score in the \CRnn \protect\subref{subfig:NN_CR}  and \SRnn \protect\subref{subfig:NN_SR} regions after performing the profile likelihood fit. Data is represented by solid circles, and the associated vertical error bar represents the statistical uncertainty on the data. The predicted yields comprise simulated \ewkwyjj signal, backgrounds from non-prompt photons and leptons that are estimated by using data-driven methods, and backgrounds that are estimated with simulation. The hashed band represents the quadrature sum of the statistical and systematic uncertainties.}
\label{fig:nn_score_postfit}
\end{figure}

%


\subsection{Signal extraction for the differential cross-section measurement}
\label{subsec:differential_signal}

The \ewkwyjj event yields are also extracted differentially as functions of \mjj,
\ptjj, \dphijj, $\pt^l$, $m_{l\gamma}$, and \dphily using the methodology documented in Ref.~\cite{STDM-2017-27}.
This method exploits signal and control regions in a binned log-likelihood fit~\cite{Barlow:1990vc,Tanabashi:2018oca}, which are used to constrain both the shape and normalisation of the \qcdwyjj background. The data are split into four regions (\SR, \CRA, \CRB, and \CRC) by
imposing the selection criteria for $\lycentrality$ and \ngapjets defined in Table~\ref{tab:sum_SRCR} in Section~\ref{sec:reconstruction}.
The variables $\lycentrality$ and \ngapjets are uncorrelated, and are thus chosen to define these four regions. The \ewkwyjj event is produced centrally and has little
hadronic activity in between the forward and backward jets, hence each event in the \SR is required to satisfy $\ngapjets = 0$ and $\lycentrality < 0.35$.
The control regions are deficient in \ewkwyjj, and are defined by inverting the requirement on either $\lycentrality$ or \ngapjets.

The binned log likelihood is defined as:
\begin{equation*}
\ln \mathcal{L} = - \sum_{r,i}\nu_{ri}^s + \sum_{r,i} N_{ri}^\mathrm{data}\,\ln{\nu_{ri}^s},
\end{equation*}

where $r$ corresponds to one of the four regions \SR, \CRA, \CRB, and \CRC, $i$ corresponds to the bin of the kinematic observable, $ N_{ri}^\mathrm{data}$ corresponds to the data yield in bin $i$ of region $r$, and $\nu^s_{ri}$ corresponds to the prediction that relies on the $s$ sources of experimental systematic uncertainty.

The fitted number of events in region $r$ and bin $i$ is expressed as
\begin{equation}
\nu_{ri} = \mu_{\mathrm{EW},i}\,\nu_{ri}^\mathrm{EW,MC} + \nu_{ri}^\mathrm{strong} + \nu_{ri}^\mathrm{other,bkg},
\end{equation}

where $\mu_{\mathrm{EW},i}$ is the signal strength of \ewkwyjj in bin $i$, and $\nu_{ri}^\mathrm{EW,MC}$ and $\nu_{ri}^\mathrm{other,bkg}$ correspond to the \ewkwyjj
prediction and contributions from reducible background processes, respectively.
The \qcdwyjj prediction is constrained using the signal-suppressed control regions based on the following four relations:

\begin{equation}
\label{eq:strongConstraints}
\begin{array}{r@{}lcr@{}l}
\nu_{\mathrm{\CRA},i}^\mathrm{strong} &{}= b_{\mathrm{L},i}\,\nu_{\mathrm{\CRA},i}^\mathrm{strong,MC}, & ~ &
\nu_{\mathrm{\CRB},i}^\mathrm{strong} &{}= b_{\mathrm{H},i}\,\nu_{\mathrm{\CRB},i}^\mathrm{strong,MC},\\[5pt]
\nu_{\mathrm{\SR},i}^\mathrm{strong} &{}= b_{\mathrm{L},i}\,c\,\nu_{\mathrm{\SR},i}^\mathrm{strong,MC} \mathrm{, ~and} & ~ &
\nu_{\mathrm{\CRC},i}^\mathrm{strong} &{}= b_{\mathrm{H},i}\,c\,\nu_{\mathrm{\CRC},i}^\mathrm{strong,MC}.
\end{array}
\end{equation}

The parameters $b_{\mathrm{L},i}$ and $b_{\mathrm{H},i}$ are sets of bin-dependent free parameters that correspond to the $\lycentrality < 0.35$ and  $\lycentrality>0.35$ regions, respectively.
The low-\lycentrality parameter, $b_{\mathrm{L},i}$, is primarily constrained by \CRA, while $b_{\mathrm{H},i}$ is primarily constrained by \CRB.
These two sets of parameters introduce additional degrees of freedom to the predicted \qcdwyjj event yield to allow the fitted number of \qcdwyjj events to be more consistent with the observed data.
A floating parameter $c$ is used to provide a residual correction that can account for any mismodelling across \ngapjets. This configuration is less prone to fit instabilities coming from the limited number of observed events than is the two-parameter linear fit of Ref.~\cite{STDM-2017-27}.

For each bin $i$, the binned maximum-likelihood fit consists of the following free parameters: the signal strength $\mu_{\mathrm{EW},i}$, the free parameter $c$, and the $b_{\mathrm{L},i}$ and $b_{\mathrm{H},i}$ corrections to the \qcdwyjj process. This constitutes an overall number of $3\times N_\mathrm{bins}+1$ parameters that are constrained by the $4\times N_\mathrm{bins}$ measurements in data, where $N_\mathrm{bins}$ corresponds to the number of bins in the measurement of the differential observable.
The alternative predictions for \ewkwyjj and \qcdwyjj introduced in Section~\ref{sec:simulation} are used to assess the systematic uncertainties discussed in Section~\ref{sec:systs}. The statistical uncertainty arising from the limited size of the data and simulated MC samples is evaluated using the bootstrap method~\cite{ATL-PHYS-PUB-2021-011}, each with 10\,000 toy experiments. The signal extraction is repeated for each toy experiment, where the event yields in each bin of the signal and control regions are sampled according to a Poisson distribution. The mean and RMS of the extracted event yields among these toy experiments are used to define the extracted \ewkwyjj yield and associated statistical uncertainty. Each source of systematic uncertainty is varied and applied to the templates coherently in each region and propagated through the fit. The theory uncertainties due to QCD scales, PDFs, and $\alpha_S$ are evaluated following the procedures discussed in Section~\ref{sec:systs}.

Figure~\ref{fig:total_systs_extr_sig} shows the pre- and post-fit agreement between data and predicted yields as functions of \mjj in \SR, \CRA, \CRB, and \CRC.
The floating parameter $c$ ranges from 1.06 to 1.22 depending on the observable while the overall scaling factor for \qcdwyjj ($b_{\mathrm{L},i}\,c$) ranges between 1.07 and 1.39.
Good post-fit agreement between the predicted yields and data is observed in all four regions.
The pre-fit systematic uncertainties presented on the figures are estimated as documented in Section~\ref{sec:systs}. The three control regions allow for a constraint of the systematic uncertainties, in particular the \qcdwyjj modelling uncertainties.

\begin{figure}
\centering
\subfloat[]{\includegraphics[width=0.49\textwidth]{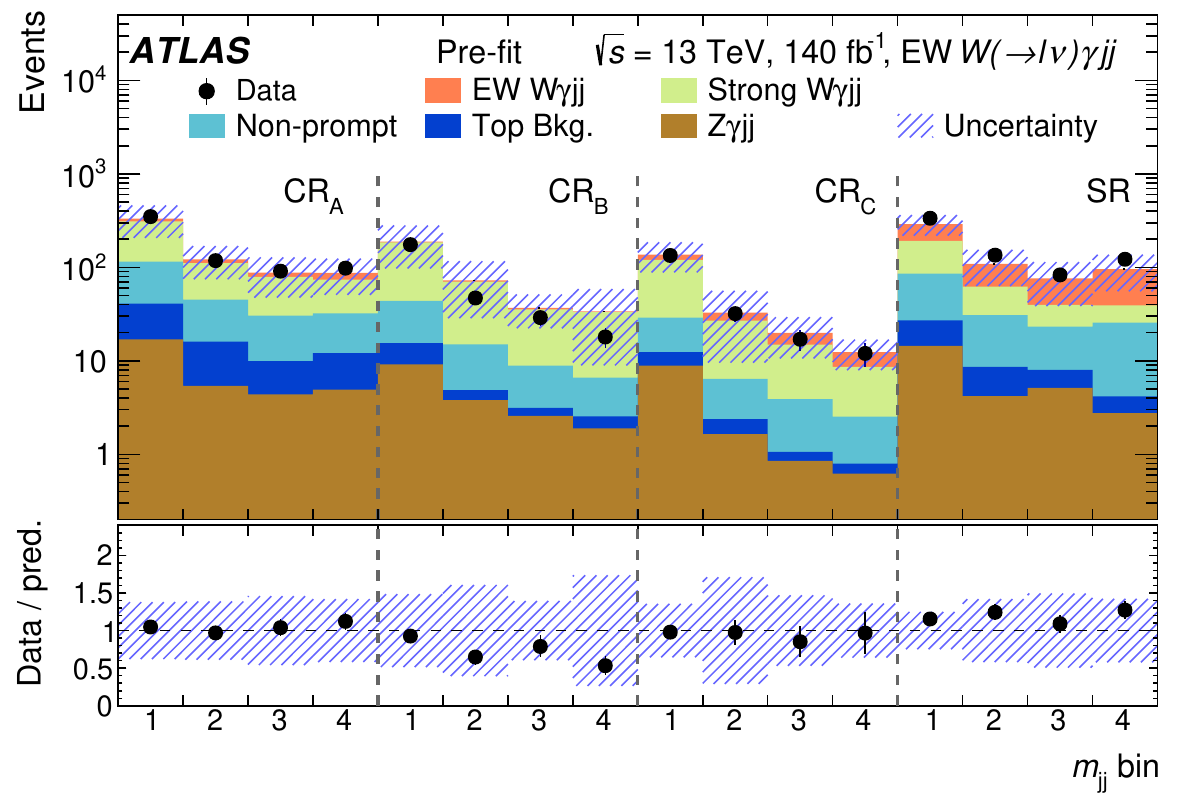}\label{subfig:stack_mjj_prefit}}
\subfloat[]{\includegraphics[width=0.49\textwidth]{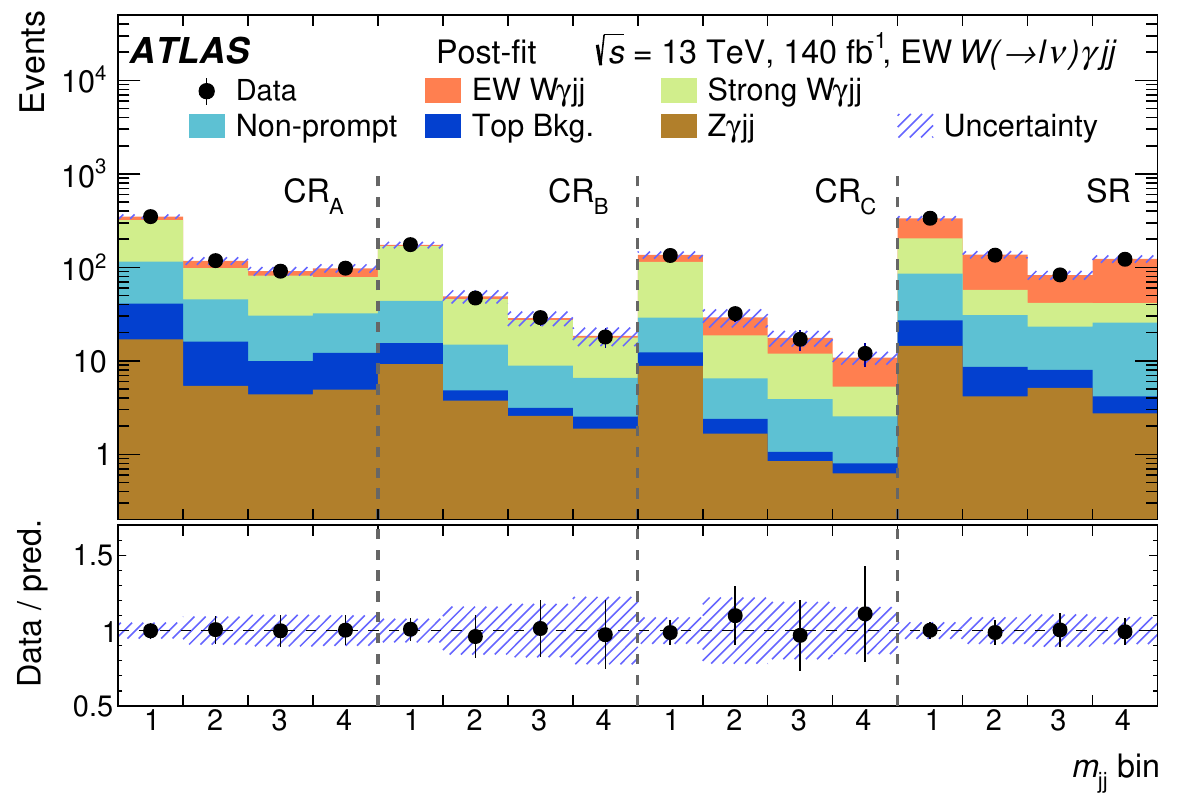}\label{subfig:stack_mjj_postfit}}
\caption{Distribution of the observed and predicted yields as a function of \mjj in the \SR, \CRA, \CRB, and \CRC regions, both \protect\subref{subfig:stack_mjj_prefit} pre-fit and \protect\subref{subfig:stack_mjj_postfit} post-fit.
Data is represented by solid circles, and the associated vertical error bar represents the statistical uncertainty of the data. The predicted yields comprise simulated \ewkwyjj signal, backgrounds from non-prompt photons and leptons that are estimated by using data-driven methods, and backgrounds that are estimated with simulation. The hashed band represents the quadrature sum of the statistical and systematic uncertainties, which are constrained in the fit. The bin edges are (1, 1.4, 1.7, 2,1, 5.3)~\TeV.}
\label{fig:total_systs_extr_sig}
\end{figure}

The differential \ewkwyjj signal extraction method is validated with two procedures.
In the first cross-check, a different choice of control regions for the \qcdwyjj process is assigned in Equation~\ref{eq:strongConstraints},
where $b_{\mathrm{L},i}$ and $b_{\mathrm{H},i}$ instead link the strong background yields at same \ngapjets\ values, and are thus
constrained at high $\lycentrality$, and the free parameter $c$ provides residual corrections considering the mismodelling across \lycentrality.
In the second method, the data-driven corrections to the \qcdwyjj contributions are derived only in either \CRA or \CRC. The extracted \ewkwyjj
event yields obtained using each of these two methods are consistent with the nominal results.


%
\section{Correction for detector effects} %
\label{sec:unfolding}
%

The fiducial and differential cross-sections at particle level are obtained by correcting the reconstruction-level \ewkwyjj event yield for the detector effects of inefficiency and resolution.

The particle-level regions are defined using all final-state stable particles with a mean lifetime of $c\tau>10$~mm.
To reduce model-dependence associated with extrapolations across phase space, the particle-level selection is defined to mimic the detector-level event selection described in Section~\ref{sec:reconstruction}.
Particle-level dressed leptons are formed by combining the four momenta of each prompt electron or muon with the prompt photons that lie within $\Delta R=0.1$ of the prompt lepton excluding electrons or muons from tau decays.
Events must contain exactly one dressed lepton with $\pt>30$~\GeV.
Leptons are required to fall within the same detector acceptance as the reconstruction level, with muons satisfying $|\eta|<2.5$ and electrons satisfying $|\eta|<2.47$ excluding the region $1.37<|\eta|<1.52$.
Events must contain at least one prompt photon with $\pt>22$~\GeV and $|\eta|<2.37$, excluding the region $1.37<|\eta|<1.52$.
At particle level, the photon isolation energy, $E_\mathrm{T}^\mathrm{iso}$ is defined as the sum of the transverse energy of stable particles within a cone of $\Delta R=0.4$ around the prompt photon, excluding the photon itself and neutrinos.
To minimise the model dependence, \SHERPA \ewkwyjj simulation is used to determine a particle-level photon isolation requirement that best mimics the detector-level behaviour; particle-level photons must satisfy $E_\mathrm{T}^\mathrm{iso}<0.2 \et^\gamma$.  %
Jets are reconstructed with the \antikt algorithm using all final-state particles as input except the dressed lepton and prompt neutrinos. Events are required to contain zero jets that fall within $\Delta R=0.4$ around a $b$-hadron, and must contain at least two jets with $\pt>50$~\GeV and $|\eta|<4.4$. The particle-level missing transverse momentum, defined as the vector sum of the transverse momenta of all non-interacting, final-state stable particles, must satisfy $\met>30$~\GeV.
These leptons, jets, and photons are then selected in a VBS topology by using the same requirements as the reconstruction-level selection, which are listed in Table~\ref{tab:fidDef}.
The particle-level phase space definition is slightly different for the fiducial and differential cross-section measurements to ensure the definitions are as close as possible to the signal extraction method at reconstruction level.

\begin{table}[t]
\centering
\caption{
Particle-level definition for the fiducial and differential \ewkwyjj  measurement.
\label{tab:fidDef}
}
\bgroup
\def\arraystretch{1.2}
\begin{tabular}{r|l}\hline
Object & Selection requirements \\
\hline
Dressed muons & $p_\mathrm{T}^{}>30$~\GeV{} and $|\eta|<2.5$ \\
Dressed electrons & $p_\mathrm{T}^{}>30$~\GeV{} and $|\eta|<2.47$ (excluding $1.37<|\eta|<1.52$) \\
Isolated photons & $\et^{\gamma}>22$~\GeV and $|\eta|<2.37$ (excluding $1.37<|\eta|<1.52$) and $E_\mathrm{T}^\mathrm{iso}<0.2 \et^\gamma$\\
Jets & At least two jets with $p_\mathrm{T}^{}>50$~\GeV{} and $|y|<4.4$, $b-$jet veto \\
Missing transverse momentum & $\met>30$~\GeV and $\mtw>30$~\GeV\\[1mm]
\hline
VBS topology &
$N_\ell = 1$, $N_\gamma \geq 1$, $|m_{\ell\gamma}-m_Z|>10$~\GeV{}\\
& $\Delta R_\text{min}(\ell,j)>0.4$,~~$\Delta  R_\text{min}(\gamma,j)>0.4$,  $\Delta R_\text{min}(\ell,\gamma)>0.4$\\
&$\Delta R_\text{min}(j_1,j_2)>0.4$,~~$\Delta  \phi_\text{min}(\met,j)>0.4$\\
& $N_\text{jets} \geq 2$,~~$p_\mathrm{T}^{j1}, p_\mathrm{T}^{j2} > 50$~\GeV{}\\
& $\mjj > 500$~\GeV{},~~$|\Delta y_{jj}|>2$\\
\hline
Fiducial measurement & VBS topology\\
Differential measurement & VBS topology~$\oplus$~($\mjj > 1000$~\GeV{}, $\Ngapjets =0$, and $\xi_{W\gamma}<0.35$)\\[1mm] \hline\hline

\end{tabular}
\egroup
\end{table}

The integrated fiducial cross-section, $\sigma_{\mathrm{EW}~W\gamma jj}^\mathrm{fid}$, is defined as :
\begin{equation}
\sigma_{\mathrm{EW}~W\gamma jj}^\mathrm{fid}=\frac{N_{\mathrm{EW}~W\gamma jj}}{L \cdot C_{\mathrm{EW}~W\gamma jj}},
\label{equ:fiducial_xsection}
\end{equation}

where $N_{\mathrm{EW}~W\gamma jj}$ is the number of extracted \ewkwyjj events after performing the fit described in Section~\ref{subsec:inclusive_signal}, $L$ is the integrated luminosity, and $C_{\mathrm{EW}~W\gamma jj}$ is the correction factor estimated by using the \ewkwyjj MC\@. The statistical uncertainty due to the limited size of data and simulation samples are propagated through Equation~\ref{equ:fiducial_xsection}, along with each source of systematic uncertainty affecting $N_{\mathrm{EW}~W\gamma jj}$ and $C_{\mathrm{EW}~W\gamma jj}$, which are treated as fully correlated.

For the differential cross-section measurement each distribution is unfolded using an iterative Bayesian method with two iterations~\cite{DAgostini:1995,Adye:2011gm}.
The binnings of the observables are optimised to ensure the relative statistical precision of the extracted \ewkwyjj event yield is similar across bins using an Asimov dataset~\cite{Cowan:2010js} formed by the $W\gamma$ and non-$W\gamma$ simulation samples.
The binning for each observable is optimised using the \ewkwyjj and \qcdwyjj simulation samples, such that a similar statistical precision is obtained for each bin of the extracted yield. An additional requirement that the bin width is not smaller than twice the resolution of the observable is enforced.
The number of iterations are determined by minimising the quadrature sum of the bias and the statistical uncertainty due to unfolding, where the bias is estimated by comparing the unfolded distribution after a certain number of iterations to the true distribution in the \ewkwyjj simulation. The value and statistical uncertainty on the fiducial \ewkwyjj cross-section are derived from the mean and root mean square (RMS) of the 10\,000 toy experiments described in Section~\ref{subsec:differential_signal}, propagated through the unfolding. These modified distributions and response matrices are then used to repeat the unfolding procedure for each toy experiment. The statistical uncertainty in each bin corresponds to the standard deviation of the unfolded results from the ensemble of 10\,000 toy experiments. Systematic uncertainties that affect both the signal extraction and the unfolding are treated as correlated and propagated through the unfolding to the final results.


%
\section{Systematic uncertainties} %
\label{sec:systs}
%

%
%
Experimental sources of systematic uncertainty arise from the reconstruction and energy calibration of objects including charged leptons, photons, jets, heavy-flavour tagging of jets, and missing transverse momentum.
These uncertainties affect both the normalisation and shape of the simulated background processes, the shape of the simulated \ewkwyjj process in the signal extraction, and the normalisation and shape of simulations used to unfold the inclusive and differential cross-sections of the \ewkwyjj process.

The lepton trigger efficiencies and lepton and photon reconstruction, identification, and isolation efficiencies in simulations are corrected using scale factors derived from data, as described in Section~\ref{sec:reconstruction}. Systematic uncertainties due to this procedure are evaluated by varying the scale factors based on their associated uncertainties~\cite{MUON-2018-03,EGAM-2018-01}. Uncertainties arising from differences between simulation and data in the reconstructed lepton (photon) momentum are evaluated by scaling and smearing the lepton (photon) transverse momentum.

Jets are calibrated using a combination of MC-based and data-driven corrections~\cite{JETM-2018-05}. Uncertainties in the measurements due to uncertainties in the jet energy scale and resolution corrections are evaluated by scaling and smearing the jet four-momentum in the simulation by the uncertainties associated with each of these corrections. Uncertainties arising from the imperfect modelling of the JVT in the simulation are estimated by varying the JVT requirement. Furthermore, uncertainties arising from differences between simulation and data in  $b$-jet tagging efficiencies are estimated by varying the associated scale factors~\cite{ATL-PHYS-PUB-2017-013}.

Uncertainties from the \met measurement are estimated by propagating the uncertainties in the transverse momenta of hard physics objects and by applying uncertainties associated with momentum scale and resolution to the track-based term~\cite{PERF-2016-07}.

Additional sources of experimental uncertainty arise due to the modelling of \pileup and the luminosity determination.
The uncertainty in the combined 2015--2018 integrated luminosity is 0.83\%~\cite{Lumi2022:ATLAS}, obtained using the LUCID-2 detector~\cite{LUCID2} for the primary luminosity measurements, complemented by measurements using the inner detector and calorimeters.

For the \ewkwyjj fiducial cross-section measurement, each source of experimental systematic uncertainty is introduced as a nuisance parameter using a Gaussian constraint in the likelihood, as described in Section~\ref{subsec:inclusive_signal}. For the differential signal extraction, 10\,000 pseudo-experiments are constructed for simulated MC samples, where each pseudo-experiment samples each simulated MC event with a unit-mean Poisson distribution. For data, in each experiment the number of data events in each bin is determined by sampling a Poisson distribution of mean equal to the nominal bin content. When evaluating the statistical significance of the systematic uncertainties, the distribution of Poisson fluctuations is the same in the nominal as that in the systematic variations.
The \ewkwyjj yield is extracted for each pseudo-experiment, and the effect of the systematic uncertainty on the yield is calculated as the mean of the distribution of differences between the varied and nominal for each pseudo-experiment.
The statistical significance of each experimental systematic uncertainty is evaluated by computing the RMS of the distribution of differences between the varied and nominal pseudo-experiments. A smoothing procedure is applied to reduce bin-to-bin statistical fluctuations such that the difference between uncertainties with or without smoothing is statistically insignificant. All experimental systematic uncertainties are treated as correlated among processes and regions.
For both the inclusive and differential cross-sections, the dominant experimental systematic uncertainties arise from the calibration of jets and modelling of \pileup conditions, resulting in approximately 4\% uncertainty in the inclusive cross-section and 5\%--9\% uncertainty in the differential measurements.

Theoretical uncertainties are accounted for from all simulated signal and background processes that affect the \ewkwyjj signal extraction.
Each source of theoretical uncertainty is estimated by repeating the signal extraction and unfolding procedure with an uncertainty variation applied and propagating this through the unfolded measurements. Theoretical uncertainties arise due to higher-order QCD corrections, the PDFs used in the simulations, the accuracy of the strong coupling constant $\alpha_\mathrm{S}$, and the choice of event generators. The effects of higher-order QCD corrections are estimated by varying each of the renormalisation ($\mu_R$) and factorisation ($\mu_F$) scales by a factor of two with $0.5\le \mu_F/\mu_R \le 2.0$.
Uncertainties due to the PDFs are evaluated by varying each Hessian eigenvector of the nominal \NNPDF[3.0nnlo] set and comparing results obtained from nominal PDFs to alternative PDFs such as \NNPDF[3.1nnlo]~\cite{NNPDF:2017mvq}, \CT[18nnlo]~\cite{CT18}, \MSHT[nnlo]~\cite{MSHT20}, and \PDFforLHC[15nnlo]~\cite{Butterworth:2015oua}.
If the difference between the nominal PDF and alternative PDF is not covered by the quadrature sum of the eigenvector changes in \NNPDF[3.0nnlo], an additional uncertainty is assigned.
The uncertainty due to $\alpha_\mathrm{S}$ is evaluated with the nominal PDF by varying the nominal $\alpha_\mathrm{S}=0.118$ by $\pm 0.001$. %
For the \ewkwyjj signal extraction, the systematic uncertainty arising from the choice of event generator for the \qcdwyjj and \ewkwyjj processes is estimated using the symmetrised envelope formed by the difference between the extracted event yield between the nominal generator, \textsc{Sherpa}, and the alternative generator, \textsc{MadGraph5+Pythia8}. For the differential cross-section measurement, the choice of event generator is extracted using the same method as with the experimental systematic uncertainties.
Treatment of the theory uncertainties is briefly discussed here. For the fiducial cross-section measurement, the QCD scale uncertainties are treated as uncorrelated between signal and control regions, because it is a conservative approach in this case as the QCD scales are varied separately instead of simultaneously in these two regions. The PDF and $\alpha_\mathrm{S}$ uncertainties are treated as correlated between regions but kept uncorrelated between different processes. Unlike the decorrelation of theory uncertainties across regions for the fiducial cross-section measurement of \ewkwyjj, this treatment of theory uncertainties is not necessary for the differential cross-section measurement for two reasons: first, the differential cross-section is measured in each bin of the observable using the fit described above, hence the theory uncertainties are decorrelated between bins of observables; and second, correlations of QCD scale variations across \ngapjets are taken into account by the floating parameter $c$ in Equation~\ref{eq:strongConstraints} in Section~\ref{subsec:differential_signal} such that only the uncertainties in $c$ are required.

Systematic uncertainties related to the unfolding procedure in the differential cross-section measurements are evaluated as follows.
First, the \ewkwyjj process simulated with the alternative generator, \textsc{MadGraph}, is used to extract the \ewkwyjj signal and unfold the data.
The difference from the nominal result is applied as a systematic uncertainty on the unfolded \ewkwyjj distributions.
Second, 10\,000 truth toy experiments are generated for the particle-level distribution according to a Gaussian distribution with mean equal to the value of the corresponding bin and RMS equal to the statistical uncertainty of the bin. The nominal response matrix is then applied to the particle-level distribution from each toy experiment to create a reconstruction-level pseudo-dataset, which is then unfolded with the nominal response matrix to produce an unfolded pseudo-dataset. The difference in the observable distribution between the truth toy experiment and the unfolded pseudo-dataset constitutes a systematic uncertainty on the unfolded result.

Table~\ref{tab:uncertainties_impact} shows the impact of the statistical and systematic uncertainties on the fiducial and differential cross-section measurement. The precision of the fiducial cross-section is limited by the statistical precision of the sample followed by the modelling uncertainties, and the largest experimental uncertainties arise from uncertainties in the jet energy scale and jet energy resolution of reconstructed jets. Figure~\ref{fig:syst_Unfold_differential} shows the fractional systematic uncertainties as functions of $m_{jj}$ and $\pt^l$. The uncertainties in the differential measurement are mostly dominated by the systematic variations arising from \qcdwyjj and \ewkwyjj modelling and the impact of the jet energy scale and resolution uncertainties, for distributions including jets.

\begin{table}[htp]
\begin{center}
\caption{Impact of uncertainties on the measured fiducial cross-section. Squared values of impacts are obtained by fixing a set of nuisance parameters of the uncertainty sources corresponding to the category to the best-fit values, then calculating the difference between the squares of the resulting uncertainty from the total uncertainty of the nominal fit.}
\begin{tabular}{lS}
\hline
\hline{Uncertainty Source} & {Fractional Uncertainty [\%]} \\
\hline
MC Statistics& 11\\
Jets& 8\\
Lepton, photon, pile-up& 8\\
\ewkwyjj modelling& 7\\
Data Statistics&6\\
\QCDwyjj modelling& 6\\
Non-prompt background & 2\\
Luminosity& 2\\
Other Background modelling& 2\\
$\met$ & 1\\
\hline
\end{tabular}
\label{tab:uncertainties_impact}
\end{center}
\end{table}

\begin{figure}[htb]
\centering
\includegraphics[width=0.450\textwidth]{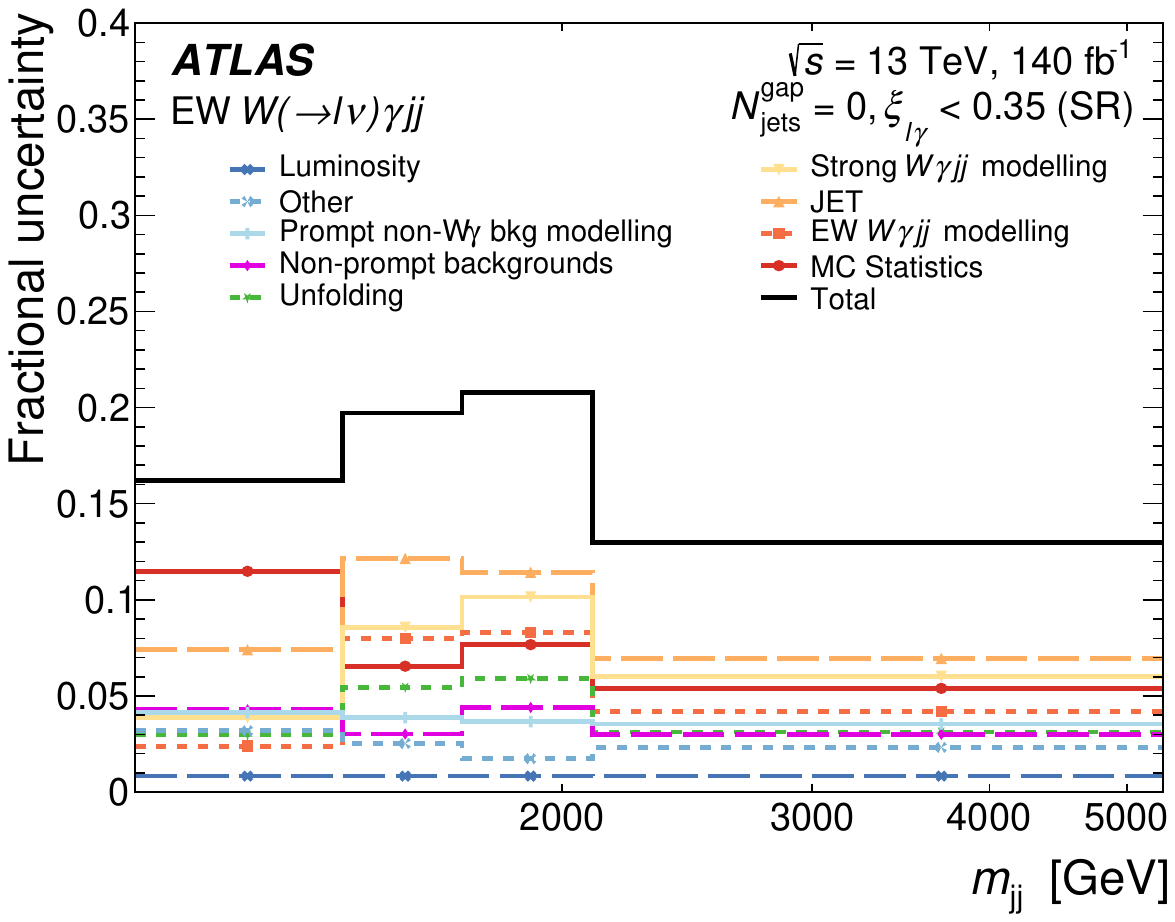}
\includegraphics[width=0.450\textwidth]{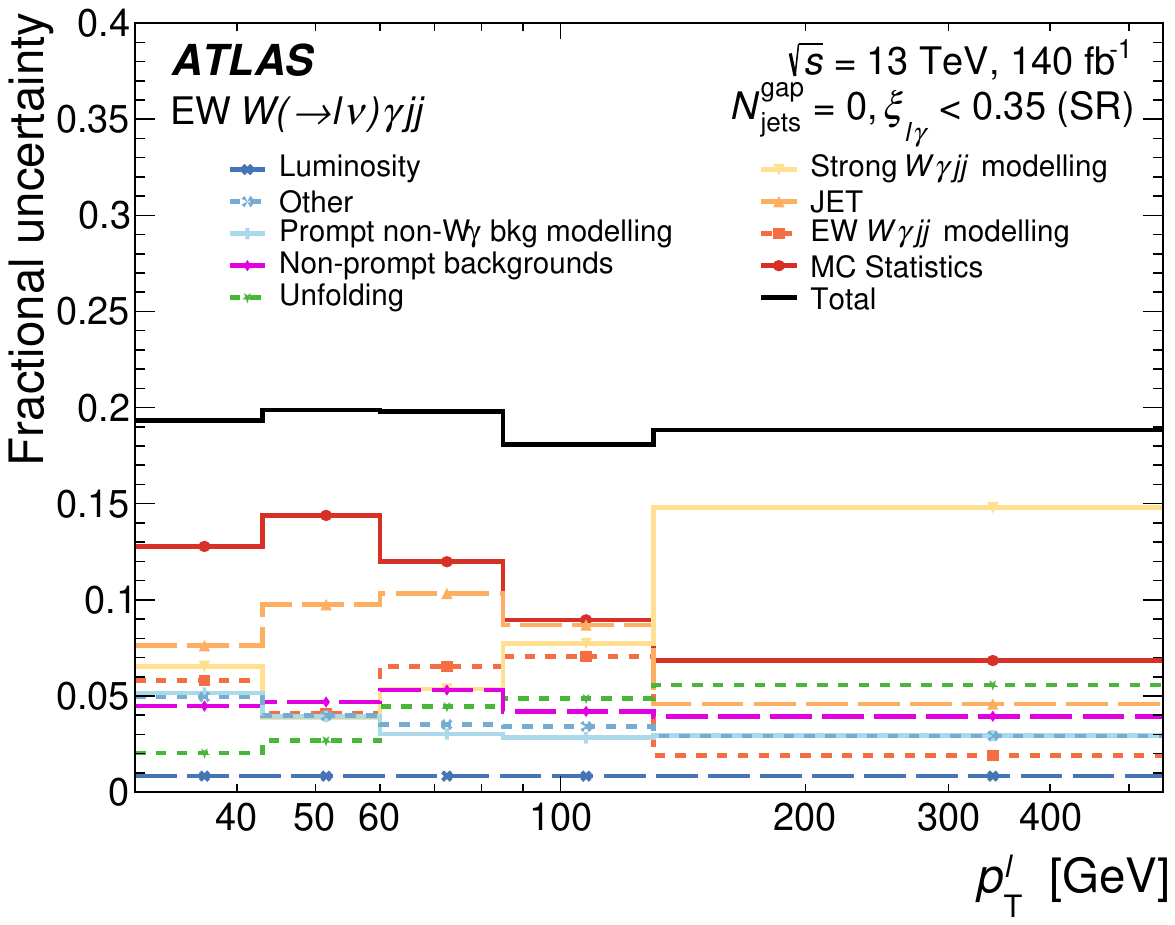}
\caption{Fractional uncertainty in the \ewkwyjj measurement as functions of $m_{jj}$ and $\pt^l$. Uncertainties are grouped in categories that are added in quadrature to give total uncertainties. Systematic uncertainties due to reconstruction of leptons, photons, pile-up, and heavy-flavour jets are grouped in the category ``Other''.}
\label{fig:syst_Unfold_differential}
\end{figure}


%
\section{Results} %
\label{sec:result}
%

\subsection{Observation and fiducial cross-section for \ewkwyjj process}

The measured signal strength, $\mu_\mathrm{EW}$, is $1.5\pm 0.5$. This includes a significant modelling uncertainty on the prediction of the fiducial cross-section due to the difference in cross-section prediction between \textsc{Sherpa} and \textsc{MadGraph5+Pythia8}, which can be seen in Figure~\ref{fig:fiducial_xs}.
The observed significance is determined using a likelihood ratio test, and is calculated to be above six standard deviations, compared with an expected significance of $6.3\sigma$.
The significance is obtained using the central value of the \textsc{Sherpa} prediction for the SM \ewkwyjj signal.
The large normalisation uncertainty on the predicted cross-section does not impact the calculation of the signal significance.

The measured \ewkwyjj fiducial cross-section in the phase space defined in Section~\ref{sec:unfolding} is determined to be $\sigma_\mathrm{EW}=13.2\pm 2.5~\mathrm{fb}$. The difference in fractional uncertainty between $\mu_\mathrm{EW}$ and $\sigma_\mathrm{EW}$ is due to a large normalisation component of the signal modelling uncertainty from the choice of event generator.
These predictions from \textsc{Sherpa} and \textsc{MadGraph5+Pythia8} are compared with the measured fiducial cross-section in Figure~\ref{fig:fiducial_xs}.
The \textsc{MadGraph5+Pythia8} prediction is in agreement with the data within uncertainties while \textsc{Sherpa} underestimates the data within two standard deviations. The theoretical uncertainties in the predicted cross-sections include systematic uncertainties from the QCD scales, PDFs, and $\alpha_S$ variations. The difference between the predicted cross-section between  \textsc{MadGraph5+Pythia8} and \textsc{Sherpa} arises due to the third parton included in the matrix element of \textsc{Sherpa}~\cite{NLOVBS}.

\begin{figure}[htb]
\centering
\includegraphics[scale=.45]{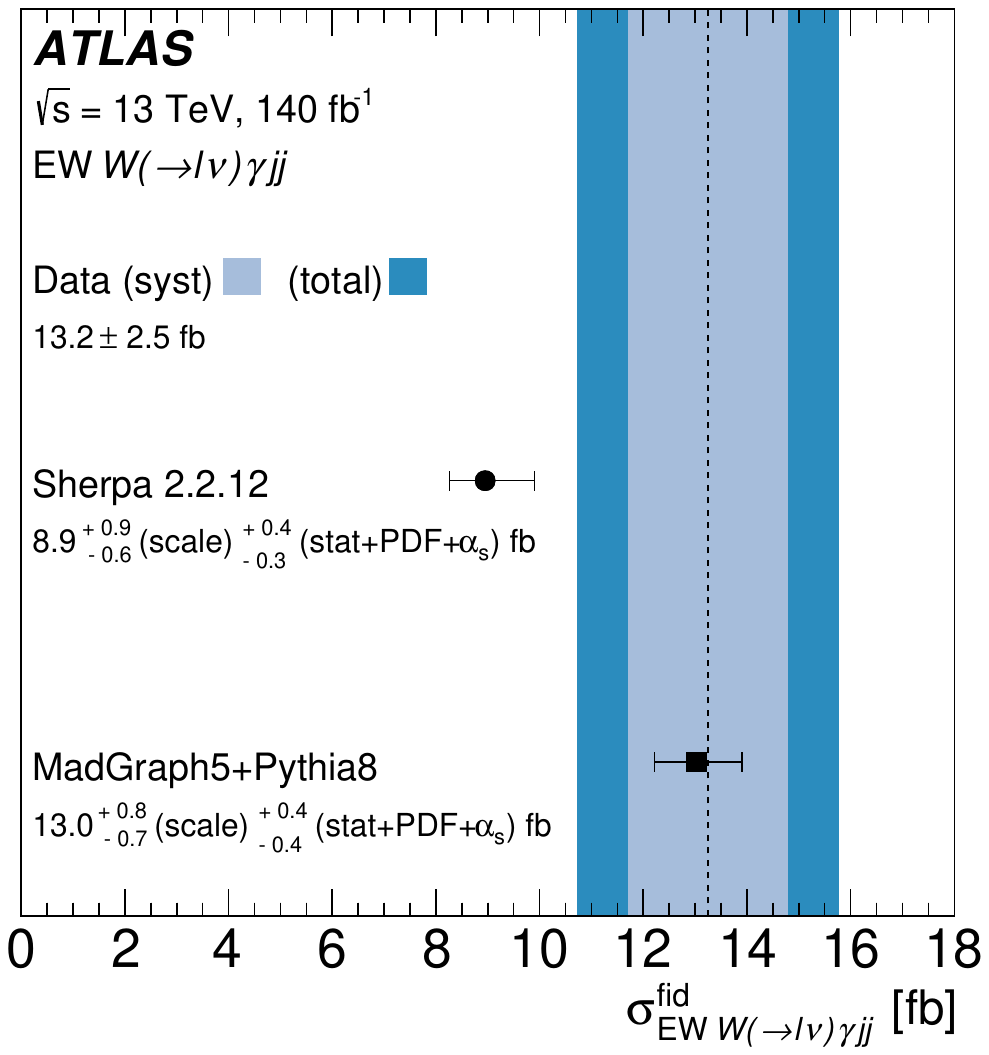}
\caption{The measured \ewkwyjj fiducial cross-section compared with the predictions of \textsc{Sherpa} and \textsc{MadGraph5+Pythia8}. The central value of the measured fiducial cross-section is represented by a dashed vertical line. The light shaded band represents the total uncertainty on the measured fiducial cross-section, while the darker shaded band represents the systematic uncertainty. Each MC prediction is represented with a solid circle, and the associated horizontal error bars correspond to the quadrature sum of statistical uncertainties and theoretical uncertainties. The impacts of various sources of systematic and theoretical modelling uncertainties in the fiducial cross-section are shown in Table~\ref{tab:uncertainties_impact}.}
\label{fig:fiducial_xs}
\end{figure}

\subsection{Differential cross-section for \ewkwyjj process}

Figure~\ref{fig:Data_Unfold_differential} shows the differential cross-sections for \ewkwyjj production as functions of $m_{jj}$, $\pt^{jj}$, $\Delta \phi_{jj}$, $\pt^l$, $m_{l\gamma}$, and $\Delta \phi_{l\gamma}$. The predictions from both \textsc{MadGraph5+Pythia8} and \textsc{Sherpa} are in agreement with the data within uncertainties. %

The presence of anomalous quartic gauge coupling can modify the distributions of these observables at the tails of $m_{jj}$, $\pt^{jj}$, $\pt^l$ or $m_{l\gamma}$. This is investigated further in Section~\ref{sec:EFTs}. Enhancement of CP violation of Higgs and gauge boson couplings in the diboson sector may distort the shape of $\Delta \phi_{jj}$ and $\Delta \phi_{l\gamma}$, thus the measurements of these two observables can be used to constrain the presence of CP-odd contributions to the \ewkwyjj process.

\begin{figure}[htbp]
\centering
\includegraphics[width=0.450\textwidth]{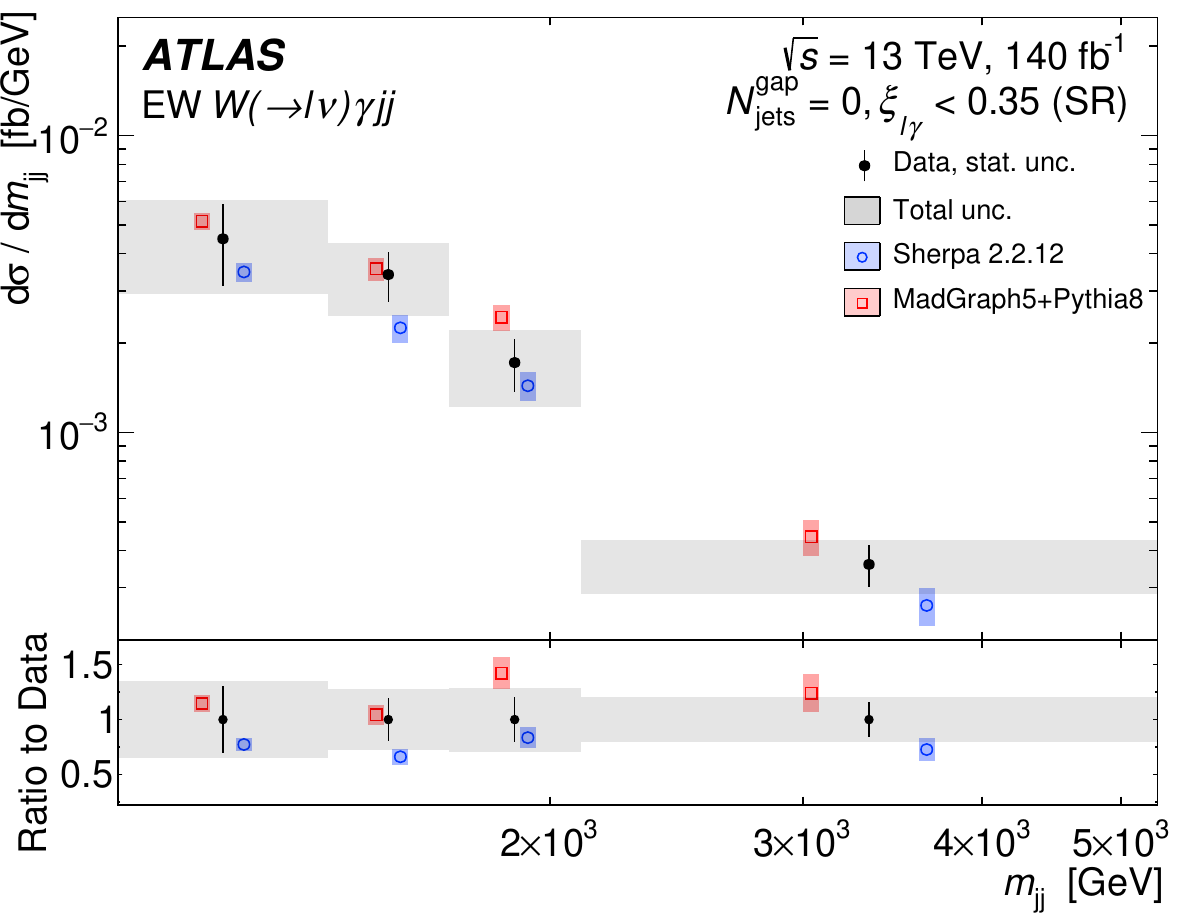}
\includegraphics[width=0.450\textwidth]{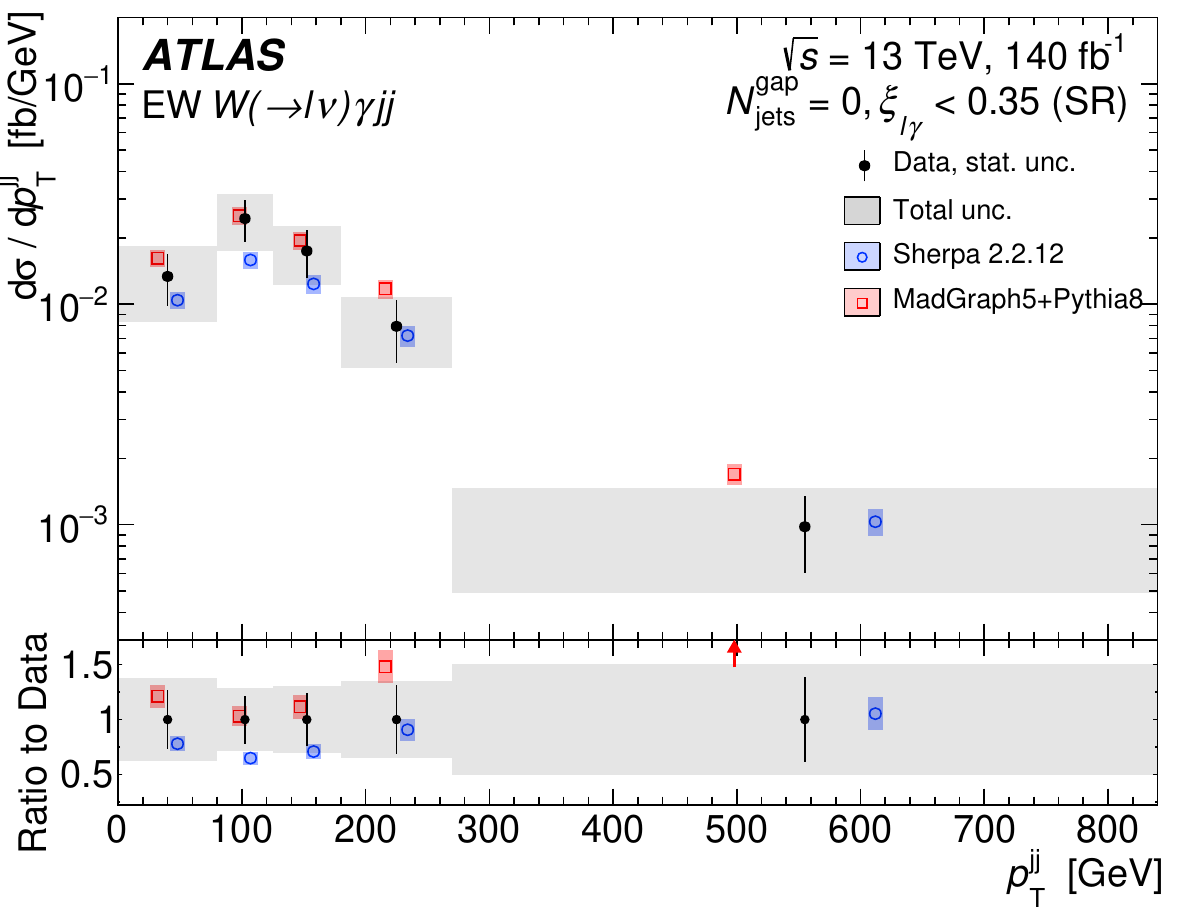}
\includegraphics[width=0.450\textwidth]{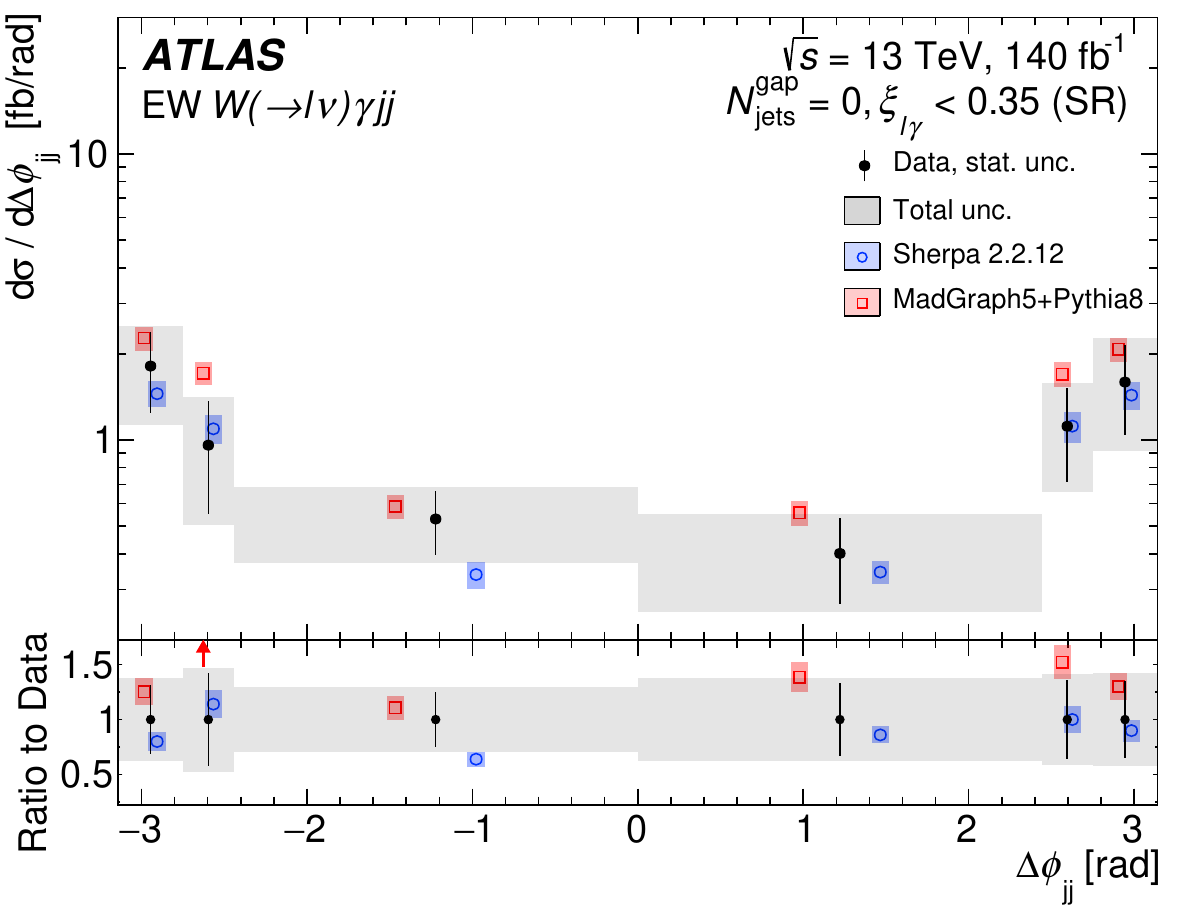}
\includegraphics[width=0.450\textwidth]{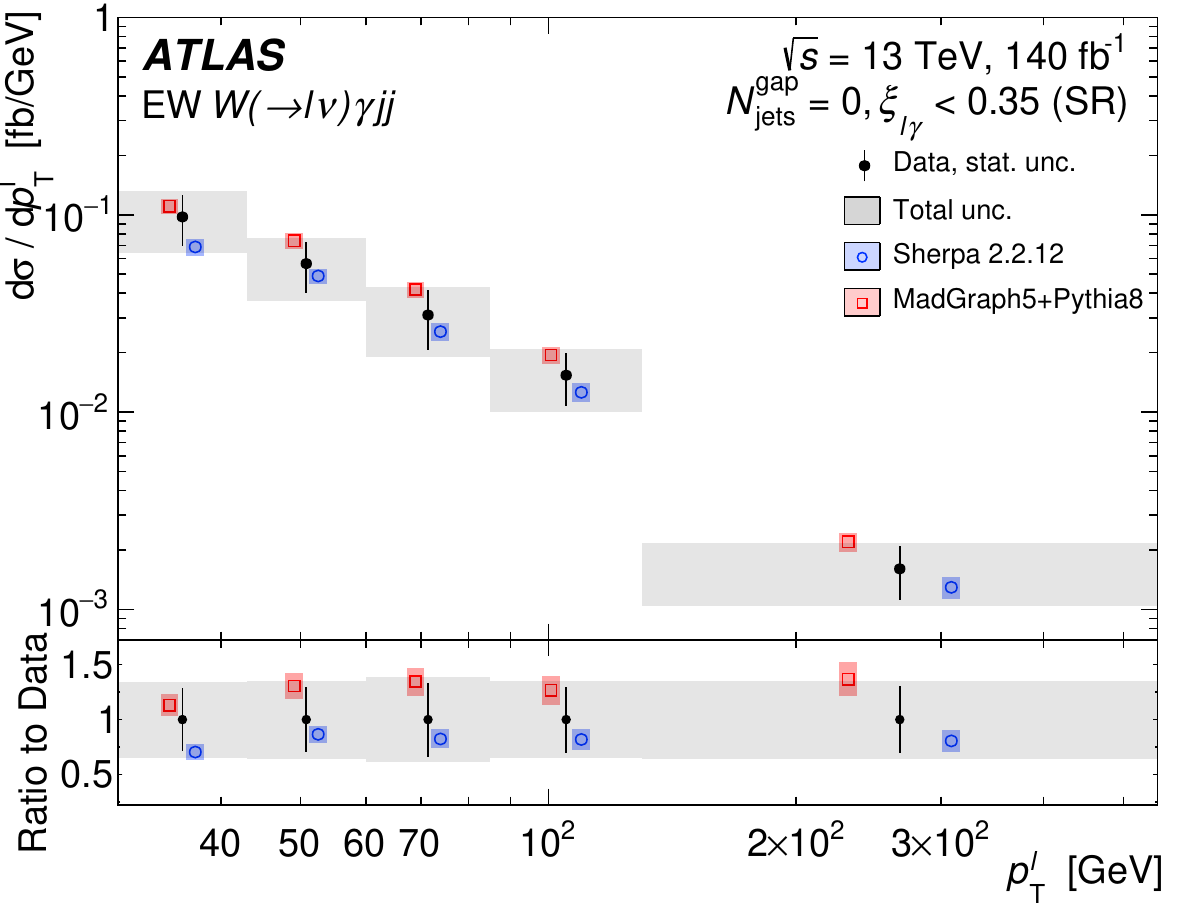}
\includegraphics[width=0.450\textwidth]{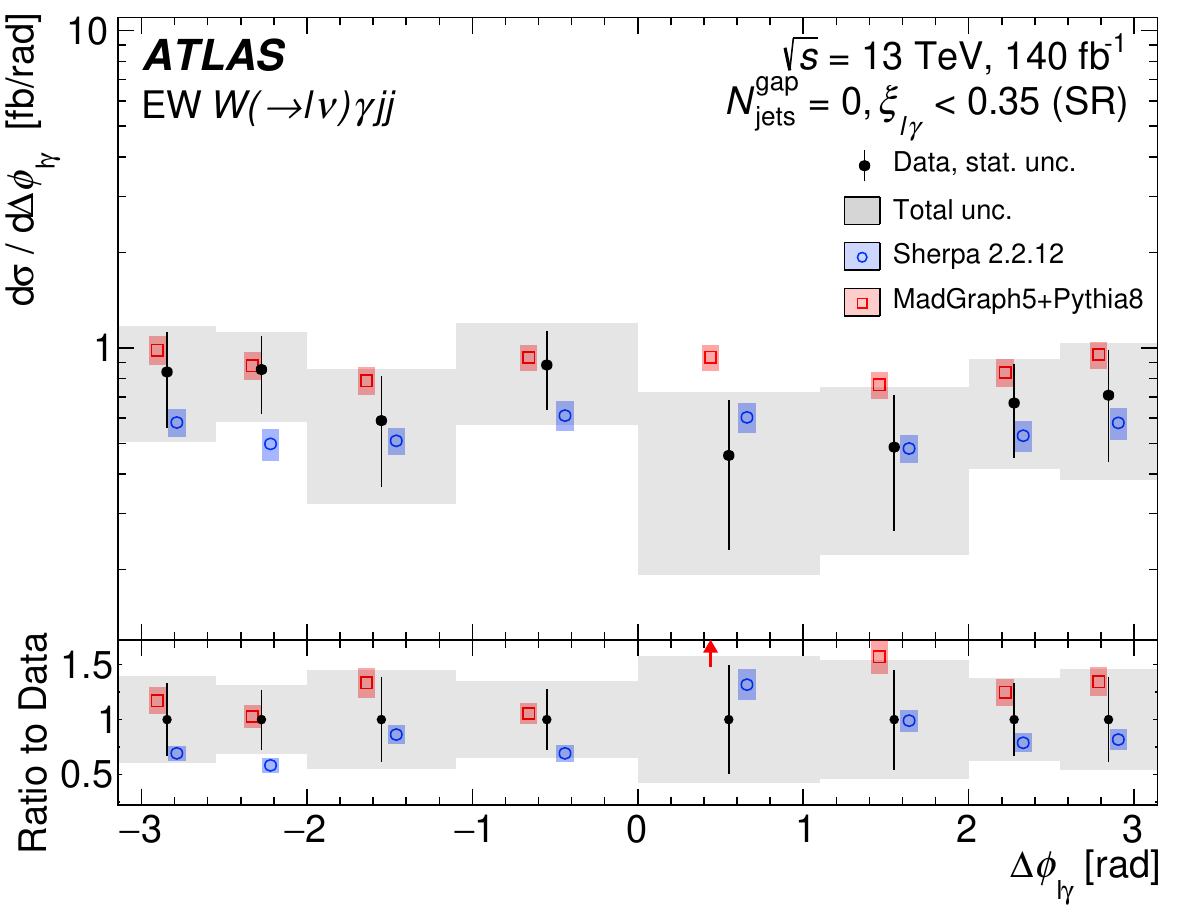}
\includegraphics[width=0.450\textwidth]{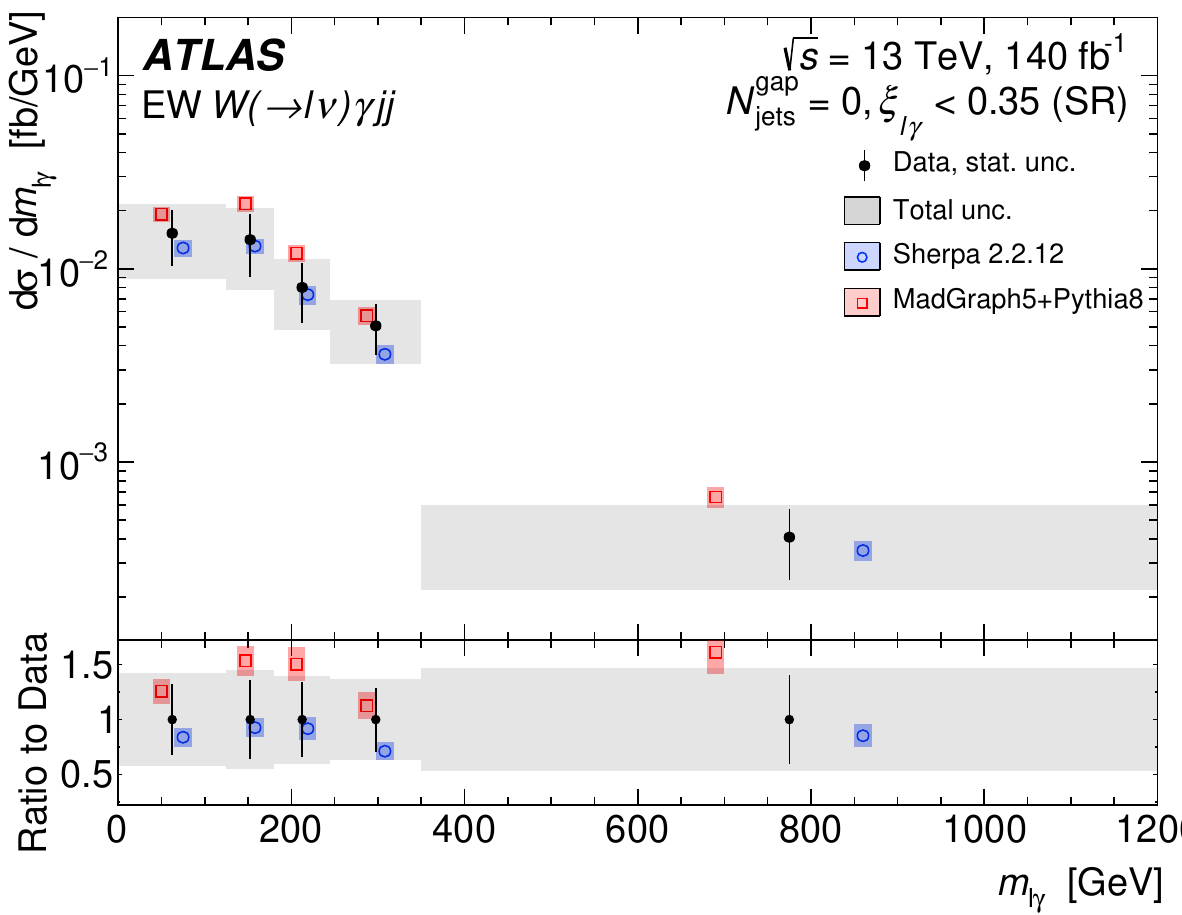}
\caption{\ewkwyjj measured differential cross-section (solid circles) as functions of $m_{jj}$, $\pt^{jj}$, $\Delta\phi_{jj}$, $\pt^l$, $m_{l\gamma}$ and $\Delta\phi_{l\gamma}$ in comparison to \textsc{Sherpa} (open circles) and \textsc{MadGraph5+Pythia8} (open squares) predictions. The vertical error bars on the predictions are determined by varying QCD scales, PDFs, $\alpha_S$ (for \textsc{Sherpa}), and MC statistical uncertainties. The vertical (shaded) error bars on the data correspond to the data statistical (total) uncertainties.}
\label{fig:Data_Unfold_differential}
\end{figure}

%


%
%
%

%
\section{EFT interpretation} %
\label{sec:EFTs}
%

%

The differential cross-section measurements presented in Section~\ref{sec:result} are used to constrain
dimension-8 (D-8) operators~\cite{PhysRevD.93.093013}. These operators are implemented in the \'{E}boli model including
twenty independent, charge-conjugation and parity conserving D-8 operators that can change the QGCs\@. The measured distributions for the six observables
are sensitive to sixteen D-8 operators and the most stringent limits from interpretations of these distributions are reported.

The effective Lagrangian, $\mathcal{L}_{\mathrm{eff}}$, including aQGCs represented by the higher dimension operators and the corresponding Wilson coefficients, is given by:

\begin{equation}
\label{eq:L_eft}
\mathcal{L}_{\mathrm{eff}} = \mathcal{L}_{\mathrm{SM}} + \sum_{j} { \frac{f_j^{(8)}}{{\Lambda}^4}} O^{(8)}_j,
\end{equation}

where $\mathcal{L}_{\mathrm{SM}}$ is the SM Lagrangian, $O^{(8)}_i$ correspond to D-8 operators with dimensionless couplings $f^{(8)}_j$ (Wilson coefficients),
and $\Lambda$ is the
energy scale of new physics. The D-8 operators are the lowest-order operators that can change QGCs without affecting the triple gauge couplings\footnote{Triple gauge couplings are represented by dimension-6 operators that affect both strong and EW production of VBS processes. These are strongly constrained by vector boson fusion processes.}. It is assumed that the  D-6 operators do not affect the \ewkwyjj or \qcdwyjj processes.
These D-8 operators can be classified into two groups: mixed-scalar operators ($O_{M0,1,2,3,4,5,7}$), consisting of two covariant derivatives of the Higgs field and two field strength tensors, and tensor-type operators, consisting of four field strength tensors ($O_{T0,1,2,3,4,5,6,7}$).

Theoretical predictions for the \ewkwyjj process are constructed based on the effective Lagrangian in Equation~\ref{eq:L_eft}.
The amplitude for the \ewkwyjj process, $\mathcal{M}$, consists of the SM contribution, $\mathcal{M}_\mathrm{SM}$, and the pure D-8 part including the aQGCs, $\mathcal{M}_\mathrm{D-8}$.

The differential cross-section can be decomposed into the following three terms:
\begin{equation}
\label{eq:eft_xsection}
|\mathcal{M}|^2=|\mathcal{M}_\mathrm{SM}|^2+2Re(\mathcal{M}_\mathrm{SM}^*\mathcal{M}_\mathrm{D-8})+|\mathcal{M}_\mathrm{D-8}|^2,
\end{equation}
where the pure SM term is $|\mathcal{M}_\mathrm{SM}|^2$, $|\mathcal{M}_\mathrm{D-8}|^2$ is the pure D-8 term that scales quadratically with $f^{(8)}_j$, and the interference term between the SM and D-8 amplitudes is $2Re(\mathcal{M}_\mathrm{SM}^*\mathcal{M}_\mathrm{D-8})$, which scales linearly with $f^{(8)}_j$.
It was found that the pure D-8 term affects the differential cross-section measurements significantly more than the interference term.
The pure SM part in Equation~\ref{eq:eft_xsection} is taken to be the LO prediction from \textsc{MadGraph5+Pythia8}, described in Section~\ref{sec:simulation}. The D-8 and interference terms are generated at LO using \textsc{MadGraph5+Pythia8}, with the same PDF and parameter tunes for modelling as the SM term. QCD scale uncertainties for the EFT contributions are taken directly from the corresponding uncertainty estimate in the SM \ewkwyjj process using \textsc{MadGraph5+Pythia8}.

Limits on the D-8 operator coefficients are determined using test statistics based on the profile likelihood ratio.
The profile likelihood ratio is constructed in terms of the measured cross-section as functions of each of the six observables and the corresponding theoretical prediction parameterised in terms of Wilson coefficients. The profile-likelihood test statistics, which are assumed to follow a $\chi^2$ distribution with one degree of freedom according to Wilks' theorem~\cite{wilks1938}, allows the 95\% confidence level (CL) bounds on Wilson coefficients. The expected 95\% coverage is further validated using 1000 pseudo experiments.

The most stringent expected limit on each coefficient is obtained from either the \ptjj or $\pt^l$ distribution.
Observed and expected limits on the Wilson coefficients at 95\% CL are presented in Table~\ref{tab:eft_limits_lep_pt} and Table~\ref{tab:lep_pt_clipped_limits} with or without unitarity preservation by introducing the clipping technique described in Ref.~\cite{Almeida:2020ylr}. The observed limits are tighter than the expected limits because \textsc{MadGraph5+Pythia8} overshoots the differential measurement shown in Figure~\ref{fig:Data_Unfold_differential}.

The limits on the D-8 Wilson coefficients using the clipping technique are estimated as functions of a cut-off scale, and for values of $M_{W\gamma}$ greater than the cut-off scale the anomalous interaction contribution is set to zero. Figure~\ref{fig:eft_clipping_scan_T3_ptjj} shows a scan of the cut-off scale for the $T3$ operator when fitting the \ptjj distribution. \ptjj is found to be the most sensitive observable to the tensor-type operators while $\pt^l$ is found to be the most sensitive observable to the mixed scalar operators. The constraints on the $f_{T3}$ and $f_{T4}$ operators represent the first such limits at the LHC.

\begin{figure}[htbp]
\centering
\includegraphics[width=.5\textwidth]{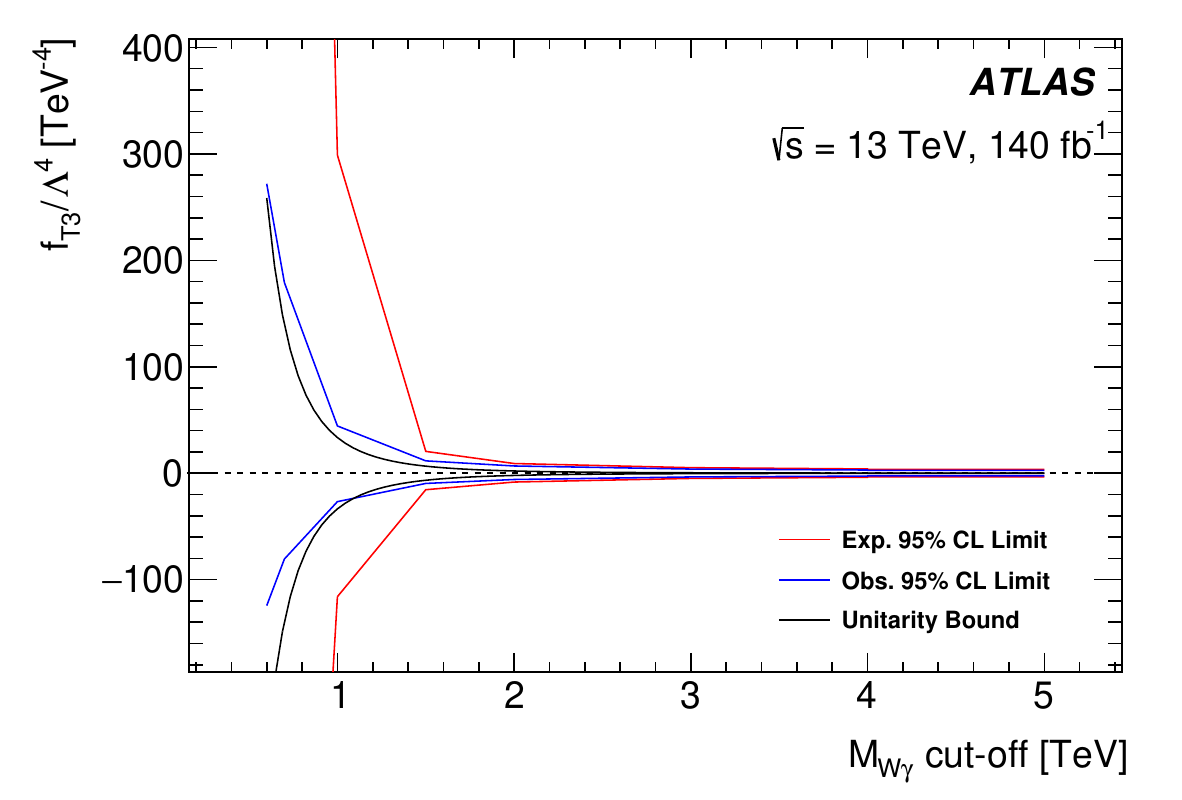}
\caption{Expected and observed 95\% CL limits of the tensor-type operator coupling $f_{T3}$ from the fit to \ptjj at different values of $m_{W\gamma}$. Bounds from partial wave unitarity constraints are also shown.}
\label{fig:eft_clipping_scan_T3_ptjj}
\end{figure}

\begin{table}[htbp]
\begin{center}
\caption{Expected and observed limits on dimension-8 operators modifying the $WW\gamma\gamma$ coupling when fitting either the $\pt^{jj}$ or $\pt^l$ distribution.}
\begin{tabular}{lccc}
\hline
Coefficients~[TeV$^{-4}$]&Observable & Expected [TeV$^{-4}$] & Observed [TeV$^{-4}$]\\
\hline

$f_{T0}/\Lambda^{4}$&$\pt^{jj}$ & { [-2.4, 2.4]} & { [-1.8, 1.8]} \\
$f_{T1}/\Lambda^{4}$&$\pt^{jj}$ & { [-1.5, 1.6]} & { [-1.1, 1.2]} \\
$f_{T2}/\Lambda^{4}$&$\pt^{jj}$ & { [-4.4, 4.7]} & { [-3.1, 3.5]} \\
$f_{T3}/\Lambda^{4}$&$\pt^{jj}$ & { [-3.3, 3.5]} & { [-2.4, 2.6]} \\
$f_{T4}/\Lambda^{4}$&$\pt^{jj}$ & { [-3.0, 3.0]} & { [-2.2, 2.2]} \\
$f_{T5}/\Lambda^{4}$&$\pt^{jj}$ & { [-1.7, 1.7]} & { [-1.2, 1.3]} \\
$f_{T6}/\Lambda^{4}$&$\pt^{jj}$ & { [-1.5, 1.5]} & { [-1.0, 1.1]} \\
$f_{T7}/\Lambda^{4}$&$\pt^{jj}$ & { [-3.8, 3.9]} & { [-2.7, 2.8]} \\

\hline
$f_{M0}/\Lambda^{4}$ & $\pt^l$& { [-28, 28]} & { [-24, 24]} \\
$f_{M1}/\Lambda^{4}$ & $\pt^l$& { [-43, 44]} & { [-37, 38]} \\
$f_{M2}/\Lambda^{4}$ & $\pt^l$& { [-10, 10]} & { [-8.6, 8.5]} \\
$f_{M3}/\Lambda^{4}$ & $\pt^l$& { [-16, 16]} & { [-13, 14]} \\
$f_{M4}/\Lambda^{4}$ & $\pt^l$& { [-18, 18]} & { [-15, 15]} \\
$f_{M5}/\Lambda^{4}$ & $\pt^l$& { [-17, 14]} & { [-14, 12]} \\
$f_{M7}/\Lambda^{4}$ & $\pt^l$& { [-78, 77]} & { [-66, 65]} \\

\hline
\end{tabular}
\label{tab:eft_limits_lep_pt}
\end{center}
\end{table}

\begin{table}[htbp]
\begin{center}

\caption{Expected and observed 95\% CL limits for specified $M_{W\gamma}$ cut-off values, where the expected limit for some operators intersects with the unitarity bounds derived from partial wave unitarity constraints. Entries with an $M_{W\gamma}$ cut-off scale of ``-" indicate that the contribution from the operators do not cross the unitarity bound over the range of the clipping scan. The fit to either the $\pt^{jj}$ or $\pt^l$ distribution is used to extract the limits.}
\begin{tabular}{lcccc}
\hline
Coefficients~[TeV$^{-4}$]& Observable&$M_{W\gamma}$ cut-off~[TeV] & Expected [TeV$^{-4}$]& Observed [TeV$^{-4}$]\\
\hline
$f_{T0}/\Lambda^{4}$& $\pt^{jj}$ & {-} & { [-2.4,2.4]} & { [-1.7,1.8]} \\
$f_{T1}/\Lambda^{4}$& $\pt^{jj}$ & {-} & { [-1.5,1.6]} & { [-1.1,1.2]} \\
$f_{T2}/\Lambda^{4}$& $\pt^{jj}$ & {-} & { [-4.4,4.7]} & { [-3.1,3.5]} \\
$f_{T3}/\Lambda^{4}$& $\pt^{jj}$ & {-} & { [-3.3,3.5]} & { [-2.4,2.6]} \\
$f_{T4}/\Lambda^{4}$& $\pt^{jj}$ & {-} & { [-3.0,3.0]} & { [-2.2,2.2]} \\
$f_{T5}/\Lambda^{4}$& $\pt^{jj}$ & {1.1} & { [-9.9,9.9]} & { [-7.5,7.5]} \\
$f_{T6}/\Lambda^{4}$& $\pt^{jj}$ & {1.3} & { [-7.4,7.6]} & { [-5.2,5.4]} \\
$f_{T7}/\Lambda^{4}$& $\pt^{jj}$ & {-} & { [-3.8,3.9]} & { [-2.7,2.8]} \\
\hline
$f_{M0}/\Lambda^{4}$& $\pt^l$  & {-} & { [-38,37]} & { [-38,37]} \\
$f_{M1}/\Lambda^{4}$& $\pt^l$  & {-} & { [-57,58]} & { [-41,42]} \\
$f_{M2}/\Lambda^{4}$& $\pt^l$  & {0.8} & { [-110,110]} & { [-88,82]} \\
$f_{M3}/\Lambda^{4}$& $\pt^l$  & {1.1} & { [-100,110]} & { [-73,77]} \\
$f_{M4}/\Lambda^{4}$& $\pt^l$  & {1.0} & { [-118,111]} & { [-89,83]} \\
$f_{M5}/\Lambda^{4}$& $\pt^l$  & {1.3} & { [-57,80]} & { [-32,77]} \\
$f_{M7}/\Lambda^{4}$& $\pt^l$  & {-} & { [-96,95]} & { [-69,68]} \\

\hline
\end{tabular}

\label{tab:lep_pt_clipped_limits}
\end{center}
\end{table}
\FloatBarrier


\section{Conclusion}
%
%

An observation of the electroweak production of $W\gamma jj$ and measurements of its fiducial and differential cross-section are reported,
using $pp$ collision data collected with the ATLAS detector at a centre-of-mass energy of $\sqrt{s}=13$~\TeV that correspond to an integrated
luminosity of 140~\ifb. This process provides a unique probe of the quartic gauge coupling via the vector boson scattering production mechanism, and measurements of \ewkwyjj production are therefore sensitive to the $WW\gamma Z$ and $WW\gamma\gamma$ quartic gauge boson couplings.

Advanced machine learning techniques are used to establish the observation of the \ewkwyjj process with a significance
of well above six standard deviations. The measured fiducial cross-section for the \ewkwyjj process is determined to be $13.2\pm 2.5$~fb, consistent with LO predictions
from \textsc{MadGraph5+Pythia8}, whereas \textsc{Sherpa} underestimates the measured cross-section.
Differential cross-sections are measured as functions of six kinematic observables in a stricter fiducial phase space.
These observables are either sensitive to the quartic gauge couplings or the CP violation structure of $WW\gamma Z$ and $WW\gamma\gamma$ couplings.
The data are corrected for detector effects of inefficiency and resolution using an iterative Bayesian unfolding method and are compared with theoretical predictions from \textsc{MadGraph5+Pythia8} and \SHERPA[2.2.12]. These unfolded distributions are described by LO predictions within uncertainties, in which \textsc{MadGraph5+Pythia8} tends to overshoot the measurement at high \mjj and $\pt^{jj}$
while \SHERPA[2.2.12] shows reasonable agreement but tends to underestimate the data across the six observables. The shape of these distributions are described reasonably well by both predictions.
These differential measurements are used to search for anomalous quartic boson interactions using dimension-8 operators in the context of an effective field theory.
The transverse momentum of the dijet system is found to be the most sensitive observable to the tensor-type operators while the charged lepton transverse momentum is found
to be the most sensitive observable to the mixed scalar operators. The first LHC constraints on $f_{T3}$ and $f_{T4}$ are presented.
\FloatBarrier


%
\section*{Acknowledgements}
%

%
%

%
%

We thank CERN for the very successful operation of the LHC and its injectors, as well as the support staff at
CERN and at our institutions worldwide without whom ATLAS could not be operated efficiently.

The crucial computing support from all WLCG partners is acknowledged gratefully, in particular from CERN, the ATLAS Tier-1 facilities at TRIUMF/SFU (Canada), NDGF (Denmark, Norway, Sweden), CC-IN2P3 (France), KIT/GridKA (Germany), INFN-CNAF (Italy), NL-T1 (Netherlands), PIC (Spain), RAL (UK) and BNL (USA), the Tier-2 facilities worldwide and large non-WLCG resource providers. Major contributors of computing resources are listed in Ref.~\cite{ATL-SOFT-PUB-2023-001}.

We gratefully acknowledge the support of ANPCyT, Argentina; YerPhI, Armenia; ARC, Australia; BMWFW and FWF, Austria; ANAS, Azerbaijan; CNPq and FAPESP, Brazil; NSERC, NRC and CFI, Canada; CERN; ANID, Chile; CAS, MOST and NSFC, China; Minciencias, Colombia; MEYS CR, Czech Republic; DNRF and DNSRC, Denmark; IN2P3-CNRS and CEA-DRF/IRFU, France; SRNSFG, Georgia; BMBF, HGF and MPG, Germany; GSRI, Greece; RGC and Hong Kong SAR, China; ISF and Benoziyo Center, Israel; INFN, Italy; MEXT and JSPS, Japan; CNRST, Morocco; NWO, Netherlands; RCN, Norway; MNiSW, Poland; FCT, Portugal; MNE/IFA, Romania; MESTD, Serbia; MSSR, Slovakia; ARRS and MIZ\v{S}, Slovenia; DSI/NRF, South Africa; MICINN, Spain; SRC and Wallenberg Foundation, Sweden; SERI, SNSF and Cantons of Bern and Geneva, Switzerland; NSTC, Taipei; TENMAK, T\"urkiye; STFC, United Kingdom; DOE and NSF, United States of America.

Individual groups and members have received support from BCKDF, CANARIE, CRC and DRAC, Canada; CERN-CZ, PRIMUS 21/SCI/017 and UNCE SCI/013, Czech Republic; COST, ERC, ERDF, Horizon 2020, ICSC-NextGenerationEU and Marie Sk{\l}odowska-Curie Actions, European Union; Investissements d'Avenir Labex, Investissements d'Avenir Idex and ANR, France; DFG and AvH Foundation, Germany; Herakleitos, Thales and Aristeia programmes co-financed by EU-ESF and the Greek NSRF, Greece; BSF-NSF and MINERVA, Israel; Norwegian Financial Mechanism 2014-2021, Norway; NCN and NAWA, Poland; La Caixa Banking Foundation, CERCA Programme Generalitat de Catalunya and PROMETEO and GenT Programmes Generalitat Valenciana, Spain; G\"{o}ran Gustafssons Stiftelse, Sweden; The Royal Society and Leverhulme Trust, United Kingdom.

In addition, individual members wish to acknowledge support from CERN: European Organization for Nuclear Research (CERN PJAS); Chile: Agencia Nacional de Investigaci\'on y Desarrollo (FONDECYT 1190886, FONDECYT 1210400, FONDECYT 1230812, FONDECYT 1230987); China: Chinese Ministry of Science and Technology (MOST-2023YFA1605700), National Natural Science Foundation of China (NSFC - 12175119, NSFC 12275265, NSFC-12075060); Czech Republic: PRIMUS Research Programme (PRIMUS/21/SCI/017); EU: H2020 European Research Council (ERC - 101002463); European Union: European Research Council (ERC - 948254, ERC 101089007), Horizon 2020 Framework Programme (MUCCA - CHIST-ERA-19-XAI-00), European Union, Future Artificial Intelligence Research (FAIR-NextGenerationEU PE00000013), Italian Center for High Performance Computing, Big Data and Quantum Computing (ICSC, NextGenerationEU); France: Agence Nationale de la Recherche (ANR-20-CE31-0013, ANR-21-CE31-0013, ANR-21-CE31-0022, ANR-22-EDIR-0002), Investissements d'Avenir Labex (ANR-11-LABX-0012); Germany: Baden-Württemberg Stiftung (BW Stiftung-Postdoc Eliteprogramme), Deutsche Forschungsgemeinschaft (DFG - 469666862, DFG - CR 312/5-2); Italy: Istituto Nazionale di Fisica Nucleare (ICSC, NextGenerationEU), Ministero dell'Università e della Ricerca (PRIN - 20223N7F8K - PNRR M4.C2.1.1); Japan: Japan Society for the Promotion of Science (JSPS KAKENHI JP21H05085, JSPS KAKENHI JP22H01227, JSPS KAKENHI JP22H04944, JSPS KAKENHI JP22KK0227); Netherlands: Netherlands Organisation for Scientific Research (NWO Veni 2020 - VI.Veni.202.179); Norway: Research Council of Norway (RCN-314472); Poland: Polish National Agency for Academic Exchange (PPN/PPO/2020/1/00002/U/00001), Polish National Science Centre (NCN 2021/42/E/ST2/00350, NCN OPUS nr 2022/47/B/ST2/03059, NCN UMO-2019/34/E/ST2/00393, UMO-2020/37/B/ST2/01043, UMO-2021/40/C/ST2/00187, UMO-2022/47/O/ST2/00148); Slovenia: Slovenian Research Agency (ARIS grant J1-3010); Spain: BBVA Foundation (LEO22-1-603), Generalitat Valenciana (Artemisa, FEDER, IDIFEDER/2018/048), Ministry of Science and Innovation (MCIN \& NextGenEU PCI2022-135018-2, MICIN \& FEDER PID2021-125273NB, RYC2019-028510-I, RYC2020-030254-I, RYC2021-031273-I, RYC2022-038164-I), PROMETEO and GenT Programmes Generalitat Valenciana (CIDEGENT/2019/023, CIDEGENT/2019/027); Sweden: Swedish Research Council (VR 2018-00482, VR 2022-03845, VR 2022-04683, VR grant 2021-03651), Knut and Alice Wallenberg Foundation (KAW 2017.0100, KAW 2018.0157, KAW 2018.0458, KAW 2019.0447, KAW 2022.0358); Switzerland: Swiss National Science Foundation (SNSF - PCEFP2\_194658); United Kingdom: Leverhulme Trust (Leverhulme Trust RPG-2020-004), Royal Society (NIF-R1-231091); United States of America: U.S. Department of Energy (ECA DE-AC02-76SF00515), Neubauer Family Foundation.

%
%


%
%
%
%
%
%
%

%
\clearpage
\appendix
\printbibliography

@Article{BernsteinPoly,
    author         = "{S., Bernstein}",
    title          = "{Proof of theorem of Weierstrass based on the calculus of probabilities}",
    journal        = "Comm. Kharkov Math. Soc.",
    volume         = "13",
    year           = "1912",
    pages          = "1-2",

}

@Article{LUCID2,
    author         = "{G. Avoni}",
    title          = "{The new LUCID-2 detector for luminosity measurement and monitoring in ATLAS}",
    journal        = "JINST",
    volume         = "13",
    year           = "2018",
    pages          = "P07017",

}

@article{wilks1938,
author = "Wilks, S. S.",
doi = "10.1214/aoms/1177732360",
fjournal = "The Annals of Mathematical Statistics",
journal = "Ann. Math. Statist.",
month = "03",
number = "1",
pages = "60--62",
publisher = "The Institute of Mathematical Statistics",
title = "The Large-Sample Distribution of the Likelihood Ratio for Testing Composite Hypotheses",
%url = "https://doi.org/10.1214/aoms/1177732360",
volume = "9",
year = "1938"
}

@Article{Eboli:2003nq,
      author         = "\'Eboli, O. J. P. and Gonzalez-Garcia, M. C. and Lietti, S.
                        M.",
      title          = "{Bosonic quartic couplings at CERN LHC}",
      journal        = "Phys. Rev. D",
      volume         = "69",
      year           = "2004",
      pages          = "095005",
      doi            = "10.1103/PhysRevD.69.095005",
      eprint         = "hep-ph/0310141",
      archivePrefix  = "arXiv",
      reportNumber   = "YITP-SB-54-03",
      SLACcitation   = "%%CITATION = HEP-PH/0310141;%%"
}

@article{PhysRevD.93.093013,
        author = "\'Eboli, O. J. P. and Gonzalez-Garcia, M. C.",
        title = "{Classifying the bosonic quartic couplings}",
        eprint = "1604.03555",
        archivePrefix = "arXiv",
        primaryClass = "hep-ph",
        reportNumber = "YITP-SB-16-09",
        doi = "10.1103/PhysRevD.93.093013",
        journal = "Phys. Rev. D",
        volume = "93",
        number = "9",
        pages = "093013",
        year = "2016"
}

@article{PhysRevD.74.073010,
  title = {Isolating vector boson scattering at the CERN LHC: Gauge cancellations and the equivalent vector boson approximation versus complete calculations},
  author = {Accomando, Elena and Ballestrero, Alessandro and Belhouari, Aissa and Maina, Ezio},
  journal = {Phys. Rev. D},
  volume = {74},
  number = {7},
  pages = {073010},
  numpages = {10},
  year = {2006},
%  month = {Oct},
  publisher = {American Physical Society},
  doi = {10.1103/PhysRevD.74.073010},
  eprint = "hep-ph/0608019",
  archivePrefix  = "arXiv",
}

@Article{CMS-SMP-21-011,
    author         = "{CMS Collaboration}",
    title          = "{Measurement of the electroweak production of \(W\gamma\) in association with two jets in proton--proton collisions at \(\sqrt{s} = 13\,\text{TeV}\)}",
    %reportNumber   = "CERN-EP-2022-223",
    journal = "Phys. Rev. D",
    volume = "108",
    pages = "032017",
    year           = "2023",
    doi            = "10.1103/PhysRevD.108.032017",
    eprint         = "2212.12592",
    archivePrefix  = "arXiv",
    primaryClass   = "hep-ex",
}

@article{Barlow:1990vc,
    author = "Barlow, Roger",
    title = "{Extended maximum likelihood}",
    reportNumber = "M-C-EXP-90-1",
    doi = "10.1016/0168-9002(90)91334-8",
    journal = "Nucl. Instrum. Meth. A",
    volume = "297",
    pages = "496--506",
    year = "1990"
}

@Article{Tanabashi:2018oca,
    author = "Tanabashi, M. and others (Particle Data Group)",
%    collaboration = "Particle Data Group",
    title = "{Review of Particle Physics}",
    doi = "10.1103/PhysRevD.98.030001",
    journal = "Phys. Rev. D",
    volume = "98",
    number = "3",
    pages = "030001",
    year = "2018"
}

@article{Almeida:2020ylr,
	archiveprefix = {arXiv},
	author = {Almeida, Eduardo da Silva and Éboli, O. J. P. and Gonzalez-Garcia, M. C.},
	doi = {10.1103/PhysRevD.101.113003},
	eprint = {2004.05174},
	journal = {Phys. Rev. D},
	number = {11},
	pages = {113003},
	primaryclass = {hep-ph},
	reportnumber = {YITP-SB-2020-8},
	title = {{Unitarity constraints on anomalous quartic couplings}},
	volume = {101},
	year = {2020}}

@Article{AdamOptimizer,
    author         = "P. Kingma, Diederik and Ba, Jimmy",
    title          = "{Adam: A Method for Stochastic Optimization}",
    %reportNumber   = "CERN-EP-2022-223",
    year           = "2014",
    eprint         = "1412.6980",
    archivePrefix  = "arXiv",
    primaryClass   = "hep-ex",
}

@inproceedings{leaky_relu,
  title="{Rectifier Nonlinearities Improve Neural Network Acoustic Models}",
  author="L. Maas, Andrew",
  year="2013",
  volume="28",
  number="3",
  booktitle="Proceedings of the 30th International Conference on Machine Learning"
  %url = "https://ai.stanford.edu/~amaas/papers/relu_hybrid_icml2013_final.pdf", 
}

@Article{Lumi2022:ATLAS,
    author         = "{ATLAS Collaboration}",
    title          = "{Luminosity determination in \(pp\) collisions at  \(\sqrt{s} = 13\,\text{TeV}\) using the ATLAS detector at the LHC}",
    year           = "2023", 
    journal = {Eur. Phys. J. C},
    volume = {83},
    number = {982},
    doi    =   {10.1140/epjc/s10052-023-11747-w},
    eprint         = "2212.09379",
    archivePrefix  = "arXiv",
    primaryClass   = "hep-ex",
}

@article{MSHT20,
  title = {Parton distributions from LHC, HERA, Tevatron and fixed target data: MSHT20 PDFs},
  author = {Bailey, S. and Cridge, T. and Harland-Lang, L. A. and Martin, A. D. and Thorne, R.S.},
  journal = {Eur. Phys. J. C},
  volume = {81},
  number = {341},
  year = {2021},
%  month = {Oct},                                                                                                                                                                   
  doi = {10.1140/epjc/s10052-021-09057-0},
  eprint = "2012.04684",
  archivePrefix  = "arXiv",
}

@Article{CT18,
    author         = "Hou, T.-J. and others",
    title          = "{Progress in the CTEQ-TEA NNLO global QCD analysis}",
    %reportNumber   = "CERN-EP-2022-223",                                                                                                                                                       
    year           = "2019",
    eprint         = "1908.11394",
    archivePrefix  = "arXiv",
    primaryClass   = "hep-ph",
}

@Article{NNPDF:2017mvq,
    author         = "{NNPDF Collaboration}",
    title          = "{Parton distributions from high-precision collider data}",
    %reportNumber   = "CERN-EP-2022-223",
   journal = {Eur. Phys. J. C},
   volume = {77},
   number = {663},
    year           = "2017",
    doi            ={10.48550/arXiv.1706.00428},
    eprint         = "1706.00428",
    archivePrefix  = "arXiv",
    primaryClass   = "hep-ex",
}

@article{HiggsCPPRL,
  title = {Determining the Structure of Higgs Couplings at the LHC},
  author = {Plehn, T. and Rainwater, D. and Zeppenfeld, D.},
  journal = {Phys. Rev. Lett.},
  volume = {88},
  number = {051801},
  year = {2002},
  doi = {10.1103/PhysRevLett.88.051801},
  eprint = "hep-ph/0105325",
  archivePrefix  = "arXiv",
  
}

@article{Higgs2jetCP,
  title = {Higgs plus two jet production via gluon fusion as a signal at the CERN LHC},
  author = {Klamke, G. and Zeppenfeld, D.},
  journal = {JHEP},
  volume = {0704},
  number = {052},
  year = {2007},
  doi = {10.1088/1126-6708/2007/04/052},
  eprint = "hep-ph/0703202",
  archivePrefix  = "arXiv",
  
}

@Article{CrystalBall,
    author         = "{Gaiser, J.}",
    title          = "{Charmonium spectroscopy from radiative decays of \(J/\Psi\) and \(\Psi^\prime\)}",
    year           = {1982},
    journal   = {SLAC-R-255},
    
}

@article{NLOVBS,
  title = {Precise predictions for same-sign W-boson scattering at the LHC},
  author = {Ballestrero, A. and others},
  journal = {Eur. Phys. J. C},
  volume = {78},
  number = {8},
  page={671},
  year = {2018},
%  month = {Oct},                                                                                                                                                                          
  doi = {10.48550/arXiv.1803.07943},
  eprint = "1803.07943",
  archivePrefix  = "arXiv",
}

@Article{Cacciari:2008gp,
     author    = "Cacciari, Matteo and Salam, Gavin P. and Soyez, Gregory",
     title     = "{The anti-\(k_{t}\) jet clustering algorithm}",
     journal   = "JHEP",
     volume    = "04",
     year      = "2008",
     pages     = "063",
     eprint    = "0802.1189",
     archivePrefix = "arXiv",
     primaryClass  =  "hep-ph",
     doi       = "10.1088/1126-6708/2008/04/063",
     SLACcitation  = "%%CITATION = 0802.1189;%%"
}

@Article{Butterworth:2015oua,
      author         = "Butterworth, Jon and others",
      title          = "{PDF4LHC recommendations for LHC Run II}",
      journal        = "J. Phys. G",
      volume         = "43",
      year           = "2016",
      pages          = "023001",
      doi            = "10.1088/0954-3899/43/2/023001",
      eprint         = "1510.03865",
      archivePrefix  = "arXiv",
      primaryClass   = "hep-ph",
      reportNumber   = "OUTP-15-17P, SMU-HEP-15-12, TIF-UNIMI-2015-14,
                        LCTS-2015-27, CERN-PH-TH-2015-249",
      SLACcitation   = "%%CITATION = ARXIV:1510.03865;%%"
}

@article{Ball:2012cx,
      author         = "{NNPDF Collaboration} and Ball, Richard D. and others",
      title          = "{Parton distributions with LHC data}",
      journal        = "Nucl. Phys. B",
      volume         = "867",
      year           = "2013",
      pages          = "244",
      doi            = "10.1016/j.nuclphysb.2012.10.003",
      eprint         = "1207.1303",
      archivePrefix  = "arXiv",
      primaryClass   = "hep-ph",
      reportNumber   = "EDINBURGH-2012-08, IFUM-FT-997, FR-PHENO-2012-014,
                        RWTH-TTK-12-25, CERN-PH-TH-2012-037, SFB-CPP-12-47\,
                        --CERN-PH-TH-2012-037",
      SLACcitation   = "%%CITATION = ARXIV:1207.1303;%%"
}

@article{Ball:2014uwa,
      author         = "{NNPDF Collaboration} and Ball, Richard D. and others",
      title          = "{Parton distributions for the LHC run II}",
      journal        = "JHEP",
      volume         = "04",
      year           = "2015",
      pages          = "040",
      doi            = "10.1007/JHEP04(2015)040",
      eprint         = "1410.8849",
      archivePrefix  = "arXiv",
      primaryClass   = "hep-ph",
      reportNumber   = "EDINBURGH-2014-15, IFUM-1034-FT, CERN-PH-TH-2013-253,
                        OUTP-14-11P, CAVENDISH-HEP-14-11",
      SLACcitation   = "%%CITATION = ARXIV:1410.8849;%%"
}

@Article{Sjostrand:2007gs,
	author    = "Sj{\"o}strand, T. and Mrenna, S. and Skands, P.",
	title     = "{A brief introduction to PYTHIA 8.1}",
	journal   = "Comput. Phys. Commun.",
	volume    = "178",
	year      = "2008",
	pages     = "852-867",
	eprint    = "0710.3820",
	archivePrefix = "arXiv",
	primaryClass  =  "hep-ph",
	doi       = "10.1016/j.cpc.2008.01.036",
	SLACcitation  = "%%CITATION = 0710.3820;%%"
}

@Article{Lange:2001uf,
      author         = "Lange, D. J.",
      title          = "{The EvtGen particle decay simulation package}",
      booktitle      = "{Proceedings, 7th International Conference on B physics
                        at hadron machines (BEAUTY 2000)}",
      journal        = "Nucl. Instrum. Meth. A",
      volume         = "462",
      year           = "2001",
      pages          = "152",
      doi            = "10.1016/S0168-9002(01)00089-4",
      SLACcitation   = "%%CITATION = NUIMA,A462,152;%%"
}

@Article{Alwall:2014hca,
      author         = "Alwall, J. and Frederix, R. and Frixione, S. and Hirschi,
                        V. and Maltoni, F. and Mattelaer, O. and Shao, H. -S. and
                        Stelzer, T. and Torrielli, P. and Zaro, M.",
      title          = "{The automated computation of tree-level and
                        next-to-leading order differential cross sections, and
                        their matching to parton shower simulations}",
      journal        = "JHEP",
      volume         = "07",
      year           = "2014",
      pages          = "079",
      doi            = "10.1007/JHEP07(2014)079",
      eprint         = "1405.0301",
      archivePrefix  = "arXiv",
      primaryClass   = "hep-ph",
      reportNumber   = "CERN-PH-TH-2014-064, CP3-14-18, LPN14-066, MCNET-14-09,
                        ZU-TH-14-14",
      SLACcitation   = "%%CITATION = ARXIV:1405.0301;%%"
}

@Article{Hoeche:2009rj,
      author         = "H{\"o}che, Stefan and Krauss, Frank and Schumann, Steffen and Siegert, Frank",
      title          = "{QCD matrix elements and truncated showers}",
      journal        = "JHEP",
      volume         = "05",
      pages          = "053",
      doi            = "10.1088/1126-6708/2009/05/053",
      year           = "2009",
      eprint         = "0903.1219",
      archivePrefix  = "arXiv",
      primaryClass   = "hep-ph",
}

@Article{Gleisberg:2008fv,
      author         = "Gleisberg, Tanju and H{\"o}che, Stefan",
      title          = "{Comix, a new matrix element generator}",
      journal        = "JHEP",
      volume         = "12",
      pages          = "039",
      doi            = "10.1088/1126-6708/2008/12/039",
      year           = "2008",
      eprint         = "0808.3674",
      archivePrefix  = "arXiv",
      primaryClass   = "hep-ph",
}

@Article{Schumann:2007mg,
      author         = "Schumann, Steffen and Krauss, Frank",
      title          = "{A parton shower algorithm based on Catani--Seymour dipole factorisation}",
      journal        = "JHEP",
      volume         = "03",
      pages          = "038",
      doi            = "10.1088/1126-6708/2008/03/038",
      year           = "2008",
      eprint         = "0709.1027",
      archivePrefix  = "arXiv",
      primaryClass   = "hep-ph",
}

@article{Hoeche:2012yf,
      author         = "H{\"o}che, Stefan and Krauss, Frank and Sch{\"o}nherr, Marek and
                        Siegert, Frank",
      title          = "{QCD matrix elements + parton showers. The NLO case}",
      journal        = "JHEP",
      volume         = "04",
      year           = "2013",
      pages          = "027",
      doi            = "10.1007/JHEP04(2013)027",
      eprint         = "1207.5030",
      archivePrefix  = "arXiv",
      primaryClass   = "hep-ph",
      reportNumber   = "SLAC-PUB-15191, IPPP-12-52, DCPT-12-104, LPN12-081,
                        FR-PHENO-2012-017, MCNET-12-09, --FR-PHENO-2012-017",
      SLACcitation   = "%%CITATION = ARXIV:1207.5030;%%"
}

@article{Bothmann:2019yzt,
      author = "Bothmann, Enrico and others",
      title = "{Event generation with Sherpa 2.2}",
      journal = "SciPost Phys.",
      volume = "7",
      year = "2019",
      number = "3",
      pages = "034",
      doi = "10.21468/SciPostPhys.7.3.034",
      reportNumber = "FERMILAB-PUB-19-218-T, SLAC-PUB-17433, IPPP/19/42, MCNET-19-11",
      eprint = "1905.09127",
      archivePrefix = "arXiv",
      primaryClass = "hep-ph",
}

@article{Hoeche:2011fd,
      author         = "H{\"o}che, Stefan and Krauss, Frank and Sch{\"o}nherr, Marek and
                        Siegert, Frank",
      title          = "{A critical appraisal of NLO+PS matching methods}",
      journal        = "JHEP",
      volume         = "09",
      year           = "2012",
      pages          = "049",
      doi            = "10.1007/JHEP09(2012)049",
      eprint         = "1111.1220",
      archivePrefix  = "arXiv",
      primaryClass   = "hep-ph",
      reportNumber   = "SLAC-PUB-14661, IPPP-11-67, DCPT-11-134, LPN11-58,
                        FR-PHENO-2011-019, MCNET-11-24",
      SLACcitation   = "%%CITATION = ARXIV:1111.1220;%%"
}

@article{Catani:2001cc,
      author         = "Catani, S. and Krauss, F. and Webber, B. R. and Kuhn, R.",
      title          = "{QCD Matrix Elements + Parton Showers}",
      journal        = "JHEP",
      volume         = "11",
      year           = "2001",
      pages          = "063",
      doi            = "10.1088/1126-6708/2001/11/063",
      eprint         = "hep-ph/0109231",
      archivePrefix  = "arXiv",
      reportNumber   = "CERN-TH-2000-367, CAVENDISH-HEP-00-03",
      SLACcitation   = "%%CITATION = HEP-PH/0109231;%%"
}

@article{Agostinelli:2002hh,
      author         = "Agostinelli, S. and others",
      title          = "{\textsc{Geant4} -- a simulation toolkit}",
      journal        = "Nucl. Instrum. Meth. A",
      volume         = "506",
      year           = "2003",
      pages          = "250",
      doi            = "10.1016/S0168-9002(03)01368-8",
      reportNumber   = "SLAC-PUB-9350, FERMILAB-PUB-03-339",
      SLACcitation   = "%%CITATION = NUIMA,A506,250;%%"
}

@Article{Cowan:2010js,
   Author = {Cowan, Glen and Cranmer, Kyle and Gross, Eilam and Vitells, Ofer},
   Title = {{Asymptotic formulae for likelihood-based tests of new physics}},
   Journal = {Eur. Phys. J. C},
   Volume = {71},
   Year = {2011},
   Pages = {1554},
   doi = "10.1140/epjc/s10052-011-1554-0",
   Eprint = "1007.1727",
   Archiveprefix = {arXiv},
   Primaryclass = {physics.data-an},
   related = "Cowan:2010js-err",
   relatedstring  = "Erratum:",
}

@Inproceedings{Adye:2011gm,
    author = "Adye, Tim",
    title = "{Unfolding algorithms and tests using RooUnfold}",
    booktitle = "{Proceedings, 2011 Workshop on Statistical Issues Related to Discovery Claims in Search Experiments and Unfolding (PHYSTAT 2011)}",
    venue = "CERN, Geneva, Switzerland",
    eventdate = "2011-01-17/2011-01-20",
    pages = "313-318",
    eprint = "1105.1160",
    archivePrefix = "arXiv",
    primaryClass = "physics.data-an",
    doi = "10.5170/CERN-2011-006.313",
}

@article{DAgostini:1995,
  author = "Giulio D'Agostini",
  title = "A multidimensional unfolding method based on Bayes' theorem",
  journal = "Nucl. Instrum. Meth. A",
  volume = "362",
  number = "2",
  pages = "487 - 498",
  year = "1995",
  issn = "0168-9002",
  doi = "10.1016/0168-9002(95)00274-X",
}

@Article{PERF-2007-01,
    author         = "{ATLAS Collaboration}",
    title          = "{The ATLAS Experiment at the CERN Large Hadron Collider}",
    journal        = "JINST",
    volume         = "3",
    year           = "2008",
    pages          = "S08003",
    doi            = "10.1088/1748-0221/3/08/S08003",
    primaryClass   = "hep-ex",
}

@Article{STDM-2010-08,
    author         = "{ATLAS Collaboration}",
    title          = "{Measurement of the inclusive isolated prompt photon cross section in \(pp\) collisions at \(\sqrt{s} = 7\,\text{TeV}\) with the ATLAS detector}",
    journal        = "Phys. Rev. D",
    volume         = "83",
    year           = "2011",
    pages          = "052005",
    doi            = "10.1103/PhysRevD.83.052005",
    reportNumber   = "CERN-PH-EP-2010-068",
    eprint         = "1012.4389",
    archivePrefix  = "arXiv",
    primaryClass   = "hep-ex",
}

@Article{PERF-2014-03,
    author         = "{ATLAS Collaboration}",
    title          = "{Performance of pile-up mitigation techniques for jets in \(pp\) collisions at \(\sqrt{s} = 8\,\text{TeV}\) using the ATLAS detector}",
    journal        = "Eur. Phys. J. C",
    volume         = "76",
    year           = "2016",
    pages          = "581",
    doi            = "10.1140/epjc/s10052-016-4395-z",
    reportNumber   = "CERN-PH-EP-2015-206",
    eprint         = "1510.03823",
    archivePrefix  = "arXiv",
    primaryClass   = "hep-ex",
}

@Article{PERF-2015-09,
    author         = "{ATLAS Collaboration}",
    title          = "{Jet reconstruction and performance using particle flow with the ATLAS Detector}",
    journal        = "Eur. Phys. J. C",
    volume         = "77",
    year           = "2017",
    pages          = "466",
    doi            = "10.1140/epjc/s10052-017-5031-2",
    reportNumber   = "CERN-EP-2017-024",
    eprint         = "1703.10485",
    archivePrefix  = "arXiv",
    primaryClass   = "hep-ex",
}

@Article{PERF-2016-07,
    author         = "{ATLAS Collaboration}",
    title          = "{Performance of missing transverse momentum reconstruction with the ATLAS detector using proton--proton collisions at \(\sqrt{s} = 13\,\text{TeV}\)}",
    journal        = "Eur. Phys. J. C",
    volume         = "78",
    year           = "2018",
    pages          = "903",
    doi            = "10.1140/epjc/s10052-018-6288-9",
    reportNumber   = "CERN-EP-2017-274",
    eprint         = "1802.08168",
    archivePrefix  = "arXiv",
    primaryClass   = "hep-ex",
}

@Article{TRIG-2016-01,
    author         = "{ATLAS Collaboration}",
    title          = "{Performance of the ATLAS trigger system in 2015}",
    journal        = "Eur. Phys. J. C",
    volume         = "77",
    year           = "2017",
    pages          = "317",
    doi            = "10.1140/epjc/s10052-017-4852-3",
    reportNumber   = "CERN-EP-2016-241",
    eprint         = "1611.09661",
    archivePrefix  = "arXiv",
    primaryClass   = "hep-ex",
}

@Article{STDM-2017-27,
    author         = "{ATLAS Collaboration}",
    title          = "{Differential cross-section measurements for the electroweak production of dijets in association with a \(Z\) boson in proton--proton collisions at ATLAS}",
    journal        = "Eur. Phys. J. C",
    volume         = "81",
    year           = "2021",
    pages          = "163",
    doi            = "10.1140/epjc/s10052-020-08734-w",
    reportNumber   = "CERN-EP-2020-045",
    eprint         = "2006.15458",
    archivePrefix  = "arXiv",
    primaryClass   = "hep-ex",
}

@Article{DAPR-2018-01,
    author         = "{ATLAS Collaboration}",
    title          = "{ATLAS data quality operations and performance for 2015--2018 data-taking}",
    journal        = "JINST",
    volume         = "15",
    year           = "2020",
    pages          = "P04003",
    doi            = "10.1088/1748-0221/15/04/P04003",
    reportNumber   = "CERN-EP-2019-207",
    eprint         = "1911.04632",
    archivePrefix  = "arXiv",
    primaryClass   = "physics.ins-det",
}

@Article{EGAM-2018-01,
    author         = "{ATLAS Collaboration}",
    title          = "{Electron and photon performance measurements with the ATLAS detector using the 2015--2017 LHC proton--proton collision data}",
    journal        = "JINST",
    volume         = "14",
    year           = "2019",
    pages          = "P12006",
    doi            = "10.1088/1748-0221/14/12/P12006",
    reportNumber   = "CERN-EP-2019-145",
    eprint         = "1908.00005",
    archivePrefix  = "arXiv",
    primaryClass   = "hep-ex",
}

@Article{JETM-2018-05,
    author         = "{ATLAS Collaboration}",
    title          = "{Jet energy scale and resolution measured in proton--proton collisions at \(\sqrt{s} = 13\,\text{TeV}\) with the ATLAS detector}",
    journal        = "Eur. Phys. J. C",
    volume         = "81",
    year           = "2021",
    pages          = "689",
    doi            = "10.1140/epjc/s10052-021-09402-3",
    reportNumber   = "CERN-EP-2020-083",
    eprint         = "2007.02645",
    archivePrefix  = "arXiv",
    primaryClass   = "hep-ex",
}

@Article{MUON-2018-03,
    author         = "{ATLAS Collaboration}",
    title          = "{Muon reconstruction and identification efficiency in ATLAS using the full Run~2 \(pp\) collision data set at \(\sqrt{s} = 13\,\text{TeV}\)}",
    journal        = "Eur. Phys. J. C",
    volume         = "81",
    year           = "2021",
    pages          = "578",
    doi            = "10.1140/epjc/s10052-021-09233-2",
    reportNumber   = "CERN-EP-2020-199",
    eprint         = "2012.00578",
    archivePrefix  = "arXiv",
    primaryClass   = "hep-ex",
}

@Article{PIX-2018-001,
    author         = "Abbott, B. and others",
    title          = "{Production and integration of the ATLAS Insertable B-Layer}",
    journal        = "JINST",
    volume         = "13",
    year           = "2018",
    pages          = "T05008",
    doi            = "10.1088/1748-0221/13/05/T05008",
    eprint         = "1803.00844",
    archivePrefix  = "arXiv",
    primaryClass   = "physics.ins-det",
}

@Article{STDM-2018-04,
    author         = "{ATLAS Collaboration}",
    title          = "{Measurement of the \(Z(\to\ell^+\ell^-)\gamma\) production cross-section in \(pp\) collisions at \(\sqrt{s} = 13\,\text{TeV}\) with the ATLAS detector}",
    journal        = "JHEP",
    volume         = "03",
    year           = "2020",
    pages          = "054",
    doi            = "10.1007/JHEP03(2020)054",
    reportNumber   = "CERN-EP-2019-228",
    eprint         = "1911.04813",
    archivePrefix  = "arXiv",
    primaryClass   = "hep-ex",
}

@Article{TRIG-2018-01,
    author         = "{ATLAS Collaboration}",
    title          = "{Performance of the ATLAS muon triggers in Run~2}",
    journal        = "JINST",
    volume         = "15",
    year           = "2020",
    pages          = "P09015",
    doi            = "10.1088/1748-0221/15/09/p09015",
    reportNumber   = "CERN-EP-2020-031",
    eprint         = "2004.13447",
    archivePrefix  = "arXiv",
    primaryClass   = "physics.ins-det",
}

@Article{TRIG-2018-05,
    author         = "{ATLAS Collaboration}",
    title          = "{Performance of electron and photon triggers in ATLAS during LHC Run~2}",
    journal        = "Eur. Phys. J. C",
    volume         = "80",
    year           = "2020",
    pages          = "47",
    doi            = "10.1140/epjc/s10052-019-7500-2",
    reportNumber   = "CERN-EP-2019-169",
    eprint         = "1909.00761",
    archivePrefix  = "arXiv",
    primaryClass   = "hep-ex",
}

@Booklet{ATL-SOFT-PUB-2021-001,
    author         = "{ATLAS Collaboration}",
    title          = "{The ATLAS Collaboration Software and Firmware}",
    howpublished   = "{ATL-SOFT-PUB-2021-001}",
    url            = "https://cds.cern.ch/record/2767187",
    year           = "2021",
}

@Booklet{ATL-SOFT-PUB-2023-001,
    author         = "{ATLAS Collaboration}",
    title          = "{ATLAS Computing Acknowledgements}",
    howpublished   = "{ATL-SOFT-PUB-2023-001}",
    url            = "https://cds.cern.ch/record/2869272",
    year           = "2023",
}

@Report{ATLAS-TDR-19,
    author         = "{ATLAS Collaboration}",
    title          = "{ATLAS Insertable B-Layer: Technical Design Report}",
    type           = "ATLAS-TDR-19; CERN-LHCC-2010-013",
    year           = "2010",
    url            = "https://cds.cern.ch/record/1291633",
    related        = "ATLAS-TDR-19-addm",
    relatedstring  = "Addendum:",
}

@Article{CMS-SMP-19-008,
    author         = "{CMS Collaboration}",
    title          = "{Observation of electroweak production of \(W\gamma\) with two jets in proton--proton collisions at \(\sqrt{s} = 13\,\text{TeV}\)}",
    journal        = "Phys. Lett. B",
    volume         = "811",
    year           = "2020",
    pages          = "135988",
    doi            = "10.1016/j.physletb.2020.135988",
    reportNumber   = "CERN-EP-2020-143",
    eprint         = "2008.10521",
    archivePrefix  = "arXiv",
    primaryClass   = "hep-ex",
}

@Booklet{ATLAS-CONF-2015-029,
    author         = "{ATLAS Collaboration}",
    title          = "{Selection of jets produced in \(13~\text{TeV}\) proton--proton collisions with the ATLAS detector}",
    howpublished   = "{ATLAS-CONF-2015-029}",
    url            = "https://cds.cern.ch/record/2037702",
    year           = "2015",
}

@Booklet{ATL-PHYS-PUB-2014-021,
    author         = "{ATLAS Collaboration}",
    title          = "{ATLAS Pythia~8 tunes to \(7~\text{TeV}\) data}",
    howpublished   = "{ATL-PHYS-PUB-2014-021}",
    url            = "https://cds.cern.ch/record/1966419",
    year           = "2014",
}

@Booklet{ATL-PHYS-PUB-2015-026,
    author         = "{ATLAS Collaboration}",
    title          = "{Vertex Reconstruction Performance of the ATLAS Detector at \(\sqrt{s} = 13~\text{TeV}\)}",
    howpublished   = "{ATL-PHYS-PUB-2015-026}",
    url            = "https://cds.cern.ch/record/2037717",
    year           = "2015",
}

@Booklet{ATL-PHYS-PUB-2016-017,
    author         = "{ATLAS Collaboration}",
    title          = "{The Pythia~8 A3 tune description of ATLAS minimum bias and inelastic measurements incorporating the Donnachie--Landshoff diffractive model}",
    howpublished   = "{ATL-PHYS-PUB-2016-017}",
    url            = "https://cds.cern.ch/record/2206965",
    year           = "2016",
}

@Booklet{ATL-PHYS-PUB-2017-013,
    author         = "{ATLAS Collaboration}",
    title          = "{Optimisation and performance studies of the ATLAS \(b\)-tagging algorithms for the 2017-18 LHC run}",
    howpublished   = "{ATL-PHYS-PUB-2017-013}",
    url            = "https://cds.cern.ch/record/2273281",
    year           = "2017",
}

@Booklet{ATL-PHYS-PUB-2021-011,
    author         = "{ATLAS Collaboration}",
    title          = "{Evaluating statistical uncertainties and correlations using the bootstrap method}",
    howpublished   = "{ATL-PHYS-PUB-2021-011}",
    url            = "https://cds.cern.ch/record/2759945",
    year           = "2021",
}
\clearpage
 
\begin{flushleft}
\hypersetup{urlcolor=black}
{\Large The ATLAS Collaboration}

\bigskip

\AtlasOrcid[0000-0002-6665-4934]{G.~Aad}$^\textrm{\scriptsize 103}$,
\AtlasOrcid[0000-0001-7616-1554]{E.~Aakvaag}$^\textrm{\scriptsize 16}$,
\AtlasOrcid[0000-0002-5888-2734]{B.~Abbott}$^\textrm{\scriptsize 121}$,
\AtlasOrcid[0000-0002-1002-1652]{K.~Abeling}$^\textrm{\scriptsize 55}$,
\AtlasOrcid[0000-0001-5763-2760]{N.J.~Abicht}$^\textrm{\scriptsize 49}$,
\AtlasOrcid[0000-0002-8496-9294]{S.H.~Abidi}$^\textrm{\scriptsize 29}$,
\AtlasOrcid[0009-0003-6578-220X]{M.~Aboelela}$^\textrm{\scriptsize 44}$,
\AtlasOrcid[0000-0002-9987-2292]{A.~Aboulhorma}$^\textrm{\scriptsize 35e}$,
\AtlasOrcid[0000-0001-5329-6640]{H.~Abramowicz}$^\textrm{\scriptsize 152}$,
\AtlasOrcid[0000-0002-1599-2896]{H.~Abreu}$^\textrm{\scriptsize 151}$
% ...
\bigskip
\\

% ...
$^{16}$Department for Physics and Technology, University of Bergen, Bergen; Norway.\\
% ...
$^{29}$Physics Department, Brookhaven National Laboratory, Upton NY; United States of America.\\
% ...
$^{35}$$^{(a)}$Facult\'e des Sciences Ain Chock, Universit\'e Hassan II de Casablanca;$^{(b)}$Facult\'{e} des Sciences, Universit\'{e} Ibn-Tofail, K\'{e}nitra;$^{(c)}$Facult\'e des Sciences Semlalia, Universit\'e Cadi Ayyad, LPHEA-Marrakech;$^{(d)}$LPMR, Facult\'e des Sciences, Universit\'e Mohamed Premier, Oujda;$^{(e)}$Facult\'e des sciences, Universit\'e Mohammed V, Rabat;$^{(f)}$Institute of Applied Physics, Mohammed VI Polytechnic University, Ben Guerir; Morocco.\\
% ...
$^{44}$Physics Department, Southern Methodist University, Dallas TX; United States of America.\\
% ...
$^{49}$Fakult\"{a}t Physik , Technische Universit{\"a}t Dortmund, Dortmund; Germany.\\
% ...
$^{55}$II. Physikalisches Institut, Georg-August-Universit\"{a}t G\"ottingen, G\"ottingen; Germany.\\
% ...
$^{103}$CPPM, Aix-Marseille Universit\'e, CNRS/IN2P3, Marseille; France.\\
% ...
$^{121}$Homer L. Dodge Department of Physics and Astronomy, University of Oklahoma, Norman OK; United States of America.\\
% ...
$^{151}$Department of Physics, Technion, Israel Institute of Technology, Haifa; Israel.\\
$^{152}$Raymond and Beverly Sackler School of Physics and Astronomy, Tel Aviv University, Tel Aviv; Israel.\\
% ...
$^{a}$ Also Affiliated with an institute covered by a cooperation agreement with CERN.\\
$^{b}$ Also at An-Najah National University, Nablus; Palestine.\\
$^{c}$ Also at Borough of Manhattan Community College, City University of New York, New York NY; United States of America.\\
$^{d}$ Also at Center for High Energy Physics, Peking University; China.\\
$^{e}$ Also at Center for Interdisciplinary Research and Innovation (CIRI-AUTH), Thessaloniki; Greece.\\
$^{f}$ Also at Centro Studi e Ricerche Enrico Fermi; Italy.\\
% ...

\end{flushleft}


%

%
%

%
%
%
%

\end{document}